\documentclass[11pt,a4paper]{article}
\pdfoutput=1
\usepackage{jheppub}
\usepackage[T1]{fontenc}
\usepackage{verbatim}   
\usepackage{color}      
\usepackage{subfigure}  
\usepackage{multirow}
\usepackage{dsfont}
\usepackage{comment}
\usepackage{pdflscape}
\usepackage[makeroom]{cancel}
\usepackage{soul}
\usepackage{rotating}
\usepackage{feynmp} 
\usepackage{float}	

\DeclareMathOperator{\Tr}{Tr}

\newcommand{\bk}[1]{\langle #1 \rangle}
\newcommand{\kk}[1]{[ #1 ]}
\newcommand{\mi}{i}
\newcommand{\QCD}{\mathrm{QCD}}
\newcommand{\EFT}{\mathrm{(\Lambda)}}
\newcommand{\tree}{\mathrm{tree}}
\newcommand{\Gr}{\mathrm{Gr}}
\newcommand{\identity}{\text{\(\mathds{1}\)}}
\newcommand{\looptxt}{\mathrm{1-loop}}
\newcommand{\MadLoop}{{\sc\small MadLoop}}

\chardef\MyArticleWithColor=\pdfcolorstackinit page direct{0 g}

\preprint{CP3-18-35, TUM-HEP-1145/18,\begin{flushright} CERN-TH-2018-136 \end{flushright}}
\begin{document}
\title{
Constraining anomalous gluon self-interactions at the LHC: a reappraisal
}
\author[a]{Valentin Hirschi,}
\author[b]{Fabio Maltoni,}
\author[b,c]{Ioannis Tsinikos,}
\author[d]{Eleni Vryonidou}
\affiliation[a]{Institute for Theoretical Physics, ETH Z\"urich, 8093 Z\"urich, Switzerland}
\affiliation[b]{Centre for Cosmology, Particle Physics and Phenomenology (CP3),\\
	Universit\'e catholique de Louvain, B-1348 Louvain-la-Neuve, Belgium} 
\affiliation[c]{Physik Department T31, Technische Universit\"at M\"unchen, James-Franck-Str. 1, \\
D-85748 Garching, Germany}
\affiliation[d]{CERN, Theoretical Physics Department, Geneva 23 CH-1211, Switzerland}	

\abstract{Anomalous self-interactions of  non-abelian gauge fields can be described by higher dimensional operators featuring gauge-invariant combinations of the field strengths. In the case of QCD, the gluon self-interactions start to be modified at dimension six by operators of the type $GGG$, with $G$ the gluon field strength tensor, possibly leading to deviations in all observables and measurements that probe strong interactions at very small distances. In this work we consider the sensitivity to the triple gluon operator of a series of observables at the LHC in di-jet, three- and multi-jet final states and heavy-quark production. We critically re-examine the robustness of long-standing as well as more recent proposals addressing issues such as the validity of the EFT expansion and the impact of higher order QCD corrections. Our results support the conclusion that multi-jet observables can reliably bound these anomalous interactions to the level that their impact on other key observables at the LHC, involving for example top quark and Higgs production, can be safely neglected. We also highlight the potential of using previously suggested angular observables in three-jet events at the LHC to further constrain these interactions.  }

\maketitle
\section{Introduction}

With about hundred inverse femtobarn of data in total collected by the CMS and ATLAS collaborations and at least twice as much expected by the end of this year, the LHC has established a golden era for precision measurements. To fully exploit the potential for detecting deviations from the Standard Model (SM) predictions and/or constrain new physics with sensitivities that go up to the multi-TeV scales, coordinated theoretical and experimental efforts are ongoing following alternative and complementary strategies.

A possibility, which has become more and more motivated by the absence of any evidence for new particles so far, is that new states might just be heavy enough to escape production at the LHC, yet coupling with the SM strongly enough to modify the interactions among SM particles via virtual particle exchanges. Thus, the search for new physics in this scenario entails accurately measuring the strength and the structure of the couplings among the SM particles and look for anomalies in their interactions. 

A general and powerful framework to analyse and parametrise deviations in SM interactions is the so-called SM Effective Field Theory (SMEFT) \cite{Weinberg:1978kz,Buchmuller:1985jz}, where the SM is augmented by a set of higher-dimensional operators
\begin{equation} 
	\mathcal{L}_\mathrm{SMEFT}=\mathcal{L}_\mathrm{SM}+
	\sum_i\frac{C_{i}}{\Lambda^2}\mathcal{O}_{i}+\mathcal{O}(\Lambda^{-4})\,,
\label{eq:smeft}
\end{equation}
which all respect the (linearly realised) SM gauge symmetries. The main hypothesis underlying this approach is that  $\Lambda$ represents the ultimate scale up to which the EFT is valid. $\Lambda$ is taken to be larger than both $v$, the Electro-Weak Symmetry Breaking (EWSB) scale in the SM, and the characteristic energy scale $\sqrt{s}$ at which the measurement is performed, {\it  i.e.}, $v/\Lambda<1$ and $s/\Lambda^2<1$. The ambitious program of laying down the theoretical basis and of devising the best experimental analyses for interpreting in the SMEFT framework the large set of precise measurements performed at LHC  has already started. This includes testing QCD as well as EW interactions, and in particular those involving heavy states such as vector bosons, the top-quark and the Higgs boson, some of which are not very well constrained yet.
In this context, considerable progress has been recently achieved in many directions, at the conceptual as well as technical level. One of the key and challenging aspects of the EFT, is related to the fact that a "global" approach is necessary to constrain higher-dimensional operators. Typically, several (if not many) operators of very different nature affect observables in a given process at the LHC, making it difficult to extract the specific information that the process is especially meant to provide.

Among the interactions that could be modified by the existence of new physics at higher scales and are of wide relevance at hadron colliders are those involving coloured particles: they are omnipresent and can be probed over a very large range of energy scales at the LHC.  A first example is that of the four-light-quark interactions, that could be mediated by new bosons, either gauge or scalars, even at the tree level. Such effects are typically searched for in di-jet final states at very high parton-parton centre of mass energy, with bounds on single operators reaching $\sim$10 TeV (with $C_i=1$) \cite{Aaboud:2017yvp,Sirunyan:2017ygf}.
Another interesting possibility, explored in this work, is that the gluon self-interactions could be modified by the following CP-conserving dimension-6 operator
\begin{equation}
O_G=g_s f_{abc} G^{a,\mu}_\nu G^{b,\nu}_\rho G^{c,\rho}_\mu\,,
\label{eq:3G}
\end{equation}
with $G_{\mu\nu}=-\frac{\mi}{g_s}[D_\mu,D_\nu]$ and $D_\mu=\partial_\mu+\mi g_s t^a A^a_\mu$. This operator can be generated at one loop by any coloured particle interacting with the gluon field minimally (the corresponding Wilson coefficients are known for different colour representations and spin of the particles running in the loop, see  \cite{Cho:1994yu}). The analogous CP-violating one, the so-called Weinberg operator~\cite{Weinberg:1989dx} $O_{\tilde G}$, where one field strength tensor is replaced by its dual counterpart: $\tilde G^{\mu\nu} = \frac{1}{2}\epsilon^{\mu\nu\rho\sigma} G_{\rho\sigma}$ starts to receive contributions at two loops. 
In fact, the CP-violating three-gluon operator is strongly constrained by low-energy measurements, such as the neutron EDM \cite{Dekens:2013zca}. The CP-even  operator, on the other hand, plays an important role in global EFT interpretations of LHC measurements as it not only enters di-jet and multi-jet process at the tree level, but also all scattering processes that feature gluon self-interactions, such as heavy quark production, possibly accompanied by weak bosons. For example, any attempt to extract information on top-quark couplings in $t\bar{t}+X$ production as pursued in~\cite{Buckley:2015lku} requires stringent constraints on the triple-gluon operator.

For this reason, already starting during the Tevatron era, proposals and strategies have been put forth to constrain the CP-conserving operator in Eq.~\ref{eq:3G}. The first relevant observation~\cite{Simmons:1989zs} is that the $2 \to 2$ parton ($q\bar{q}\rightarrow g g$ and $g g \rightarrow g g$) amplitudes featuring \emph{a single} anomalous three-gluon interaction, do \emph{not} interfere with the corresponding SM QCD amplitudes at tree-level. The three-gluon operator therefore only contributes to di-jet production in matrix elements at order $\mathcal{O}(1/\Lambda^4)$ \cite{Dreiner:1991xi}. This fact led to considering alternative observables, in multi-jet final states~\cite{Dixon:1993xd}, in 4-jet events in electron-positron collisions \cite{Dreiner:1991xi} and in heavy quark production~\cite{Cho:1993eu,Cho:1994yu} where terms both linear and quadratic in $1/\Lambda^2$ contribute. Attempts to constrain the $O_G$ operator using the dedicated three-jet observables suggested in~\cite{Dixon:1993xd} have so far not been pursued at the LHC, while constraints from top pair production have been obtained in~\cite{Buckley:2015lku}. 

Motivated by new data made available on multi-jet measurements by the CMS collaboration performed in the context of search for black holes \cite{Sirunyan:2017anm}, Krauss, Kuttimalai and Plehn~\cite{Krauss:2016ely} recently argued that strong constraints on $O_G$ could be obtained by using high-multiplicity jet measurements at the LHC, more specifically using the particular observable
\begin{equation}
\label{STdef}
S_T=\cancel{E_T}+\sum^{N_{\textrm{jets}}}_{j=1} E_{T,j}, 
\end{equation}
where the sum runs over all jets with $p_T$ above 50 GeV as well as missing transverse energy $\cancel{E_T}$ if exceeding 50 GeV. This observable turns out to be sensitive to $O_G$ in the high-energy region of $S_T>2$ TeV, where data are available. This study finds that the sensitivity increases with the number of jets, and the measurement of the $S_T$ distributions sets a stringent constraint on the operator:
\begin{equation}
\frac{C_G}{\Lambda^2}<(5.2\, \textrm{TeV})^{-2}.
\end{equation}
Taken at face value, this strong limit would imply that, currently, the triple gluon operator should be ignored in EFT investigations of all other processes of interest at the LHC as well as in global EFT fits. However,  several questions and possible pitfalls arise concerning the sensitivity from this observable, some of which were already discussed in ref.~\cite{Krauss:2016ely}. For example, it was found that the constraint originates from the higher order contributions of the operator, {\it i.e.}, $\mathcal{O}(1/\Lambda^n)$ with $n\ge 4$, as the linear contributions stay negligible even when the number of jets exceeds two. As the limit is then based on considering  $O_G$ contributions that start at $\mathcal{O}(1/\Lambda^4)$, one must carefully assess under which hypotheses this result can be relied upon. In what follows, we shall elaborate on this potential limitation and discuss several other issues that call for a detailed study before this limit based on multi-jet data can be used in a broad context.

The goal of this work is multifold. First, we investigate the robustness of the limit of ref.~\cite{Krauss:2016ely} in light of the limited validity of the EFT expansion. More specifically, we will assess the validity of the EFT
expansion considering also higher-order contributions in $1/\Lambda^2$  to the observable considered in this particular CMS analysis. Second, we study in detail the impact of the three-gluon operator in other jet observables, including hard and well-separated jet configurations to be able to build a consistent picture of how the numerically leading contributions
appear in multi-jet events. Third, having established the reliability of the limits, we re-examine the relevance of the triple gluon operator in heavy quark production. Fourth, we determine the sensitivity of di-jet production at order $\mathcal{O}(1/\Lambda^2)$ by computing for the first time the four-parton one-loop amplitudes featuring one insertion of the $O_G$ operator. 
Finally, we follow up on the original suggestion of ref.~\cite{Dixon:1993xd} and identify regions where the linear contributions are important by making use of the different behaviour of pure QCD and $O_G$ amplitudes in three-jet events. We extend the analysis at the LHC, critically assessing the relevance of observables for which terms linear in $1/\Lambda^2$ can provide meaningful constraints. 

\section{A critical look at the bounds from multi-jet measurements}
The stringent bound set on $O_G$  in \cite{Krauss:2016ely} using the high-multiplicity jet measurements is intriguing. On the one hand, it implied that the triple gluon operator can be ignored in most SMEFT analyses that are and will be performed at the LHC. On the other hand,  the bound is to a large extent unexpected and raises a number of questions, mostly related to its robustness. We list these questions below:
\begin{itemize}
\item The experimental selection of the CMS analysis implies that the $S_T$ measurement is dominated by di-jet-like configurations, even for the high-multiplicity samples. This naturally leads us to question why the sensitivity improves with the number of jets.
\item The limits are set using data in the high-energy region, with $S_T\sim \Lambda$, which requires further analysis to ensure the EFT condition of $E<\Lambda$ is satisfied. 
\item The dominant contribution comes from the higher-order terms {\it i.e.}, $\mathcal{O}(1/\Lambda^n)$ with $n\ge 4$. One has therefore to first understand why this is the case and whether it is in general or specific for the observables considered. In addition, further investigation is needed to establish whether such terms dominate over possible dimension-8 operators, formally of the same order, which in principle could also give an important contribution to the relevant observables. 
\item The limit is set from the observable $S_T$, and it is therefore worth exploring other multi-jet observables that could be potentially used to improve this limit. For instance, configurations involving hard and well separated jets could be considered. 
\end{itemize}
In this section we scrutinise the results of \cite{Krauss:2016ely} and carefully address each of the above-mentioned points.

\label{multiplicitysection}
\subsection{$O_G$ effects and jet multiplicity}
\begin{figure}[t!]
\begin{center}
\includegraphics[width=0.7\textwidth,trim=0cm 3cm 0cm 0cm]{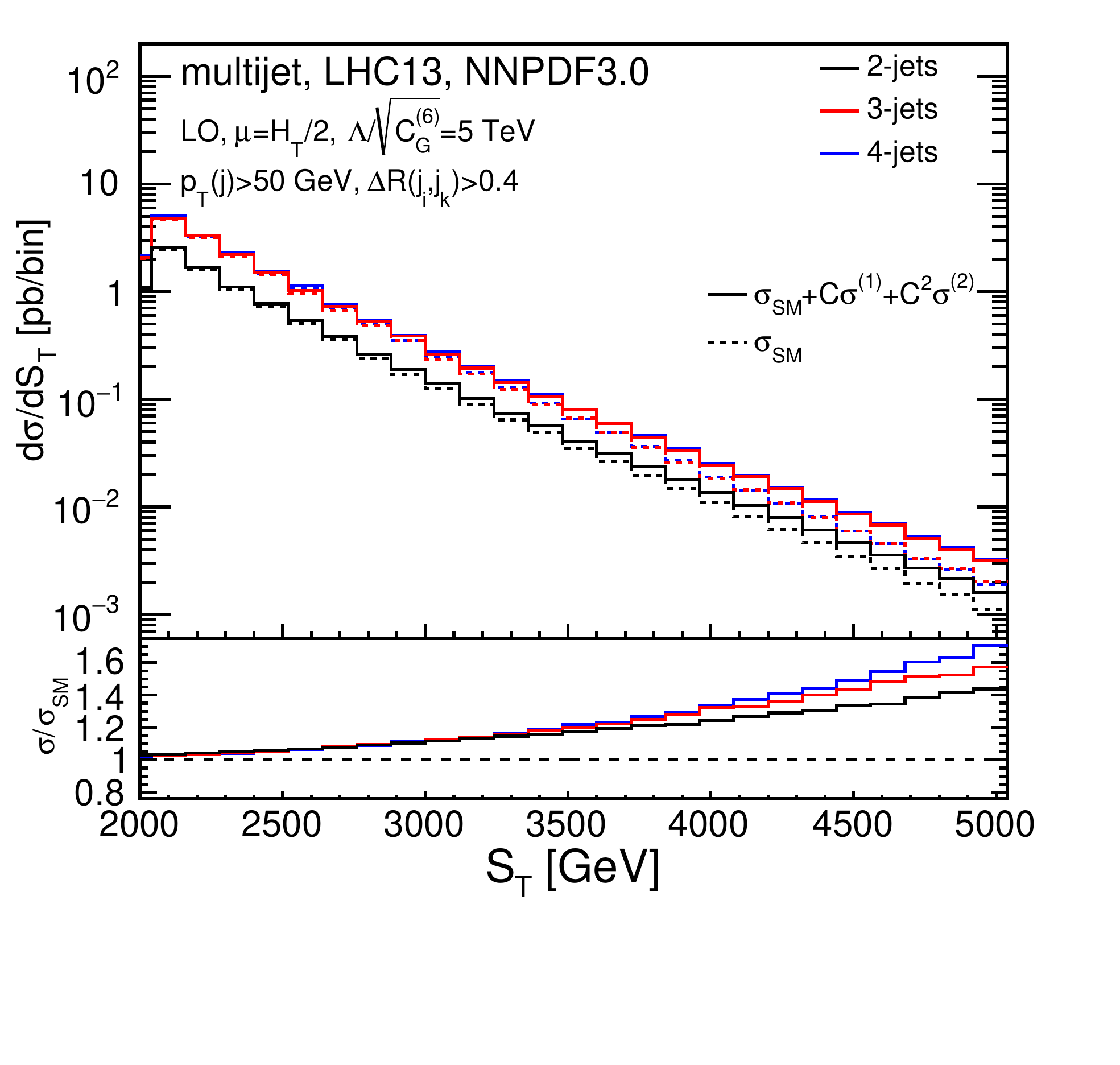}
\caption{\label{fig:noofjets}Impact of jet multiplicity on the sensitivity to $O_G$ for multi-jet production at LHC13. A maximum of one insertion of the $O_G$ operator is allowed in all of the amplitudes considered and the $\mathcal{O}(1/\Lambda^4)$ contributions are kept.}
\end{center}
\end{figure}

The first interesting observation of \cite{Krauss:2016ely} is the fact that the sensitivity of $S_T$ on $O_G$ increases with the number of jets. We have reproduced this observation in Figure~\ref{fig:noofjets},  where we show results for the production of 2, 3 and 4 partonic jets in the SM, considering amplitudes with at most one $O_G$ operator insertion and including the $\mathcal{O}(1/\Lambda^4)$ contribution from the square of these amplitudes. We find that the ratio over the SM increases with $S_T$ but also with the number of partonic jets considered. We note here that whilst \cite{Krauss:2016ely} considers predictions obtained from merging event samples featuring different partonic multiplicities, for simplicity our investigation is based on separate simulations of various multiplicities of well-isolated partonic jets. All results in this work are obtained using the {\sc MadGraph5\_aMC@NLO} ({\sc MG5\_aMC}) framework \cite{Alwall:2014hca}.

\begin{figure}[h!]
\begin{center}
\includegraphics[width=0.7\textwidth,trim=0cm 3cm 0cm 0cm]{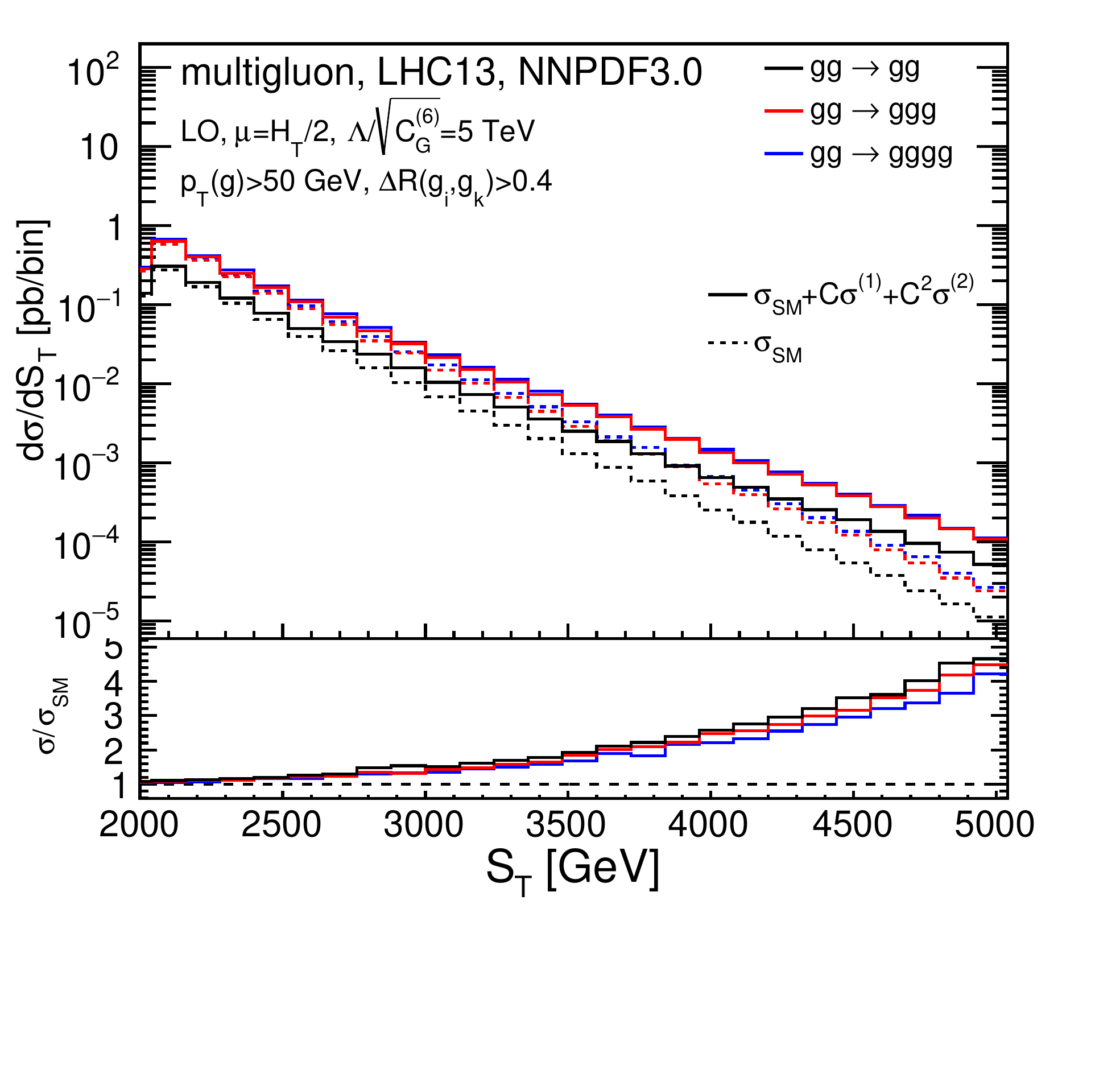}
\caption{\label{fig:gluonjets}Impact of jet multiplicity on the sensitivity to $O_G$ from only-gluon channels.}
\end{center}
\end{figure}

For the loose jet $p_T$ selection cuts of $p_T > 50$ GeV, the 3- and 4-jet final states with $S_T>2$ TeV are dominated by di-jet configurations, where two jets are hard and back-to-back, accompanied by one or two additional soft jets. One therefore wonders why the sensitivity increases with the number of jets, given that the kinematical configurations are indeed di-jet-like, and extra soft radiation cannot influence short-distance physics as the one described by the $O_G$ operator.

In order to further study the origin of this increase in sensitivity with the jet multiplicity, we investigate in Figure~\ref{fig:gluonjets} the contributions from gluon-only channels, that is $gg\to gg$, $gg\to ggg$ and $gg\to gggg$. We find that there is no increase in the sensitivity in this case, suggesting that this effect is not driven by the behaviour of the matrix elements for individual channels. This is consistent with the physical picture that long-distance (soft) radiation cannot affect phenomena at short-distance (hard).  Instead, the increase of the ratio over the SM yield when more jets are considered is mostly a due to the fact that new partonic channels open up. This is confirmed by studying the breakdown of the various contributing channels for the production of 2, 3 and 4 partons in the large-$S_T$ region, revealing that the opening of new channels receiving non-zero contributions from $O_G$ is indeed responsible for the increased sensitivity. This breakdown is presented in Tables \ref{table:2-jet}-\ref{table:4-jet}  in Appendix \ref{app:tables} for jets with $p_T>50$ GeV, along with the ratio to the corresponding SM prediction for each subprocess. At high $S_T$ the quark-initiated channels dominate due to the relative importance of the valence quark luminosity at high $x$.  Di-jet production receives no interference contributions of order  $\mathcal{O}(1/\Lambda^2)$. Contributions from the square of amplitudes for the subprocesses with two and four gluons and featuring exactly one $O_G$ operator insertion lead to terms of order $\mathcal{O}(1/\Lambda^4)$, while the $O_G$ operator cannot contribute to the four-quark process. Overall, this suppresses the $O_G$ contribution to di-jet production and make it small compared to the SM yield, especially for large Bjorken $x$'s where the quark-initiated processes dominate.
Once more final-state partons are considered, all subprocesses receive contributions from $O_G$, in particular the ones enhanced by the valence quark PDFs, and the overall ratio over the SM increases. In other words,  the increase in sensitivity with the number of jets is mostly related to the interplay between the different luminosity of the various partonic channels and their different dependence to the $O_G$ operator. In fact, based on Figure~\ref{fig:gluonjets} and Tables \ref{table:2-jet}-\ref{table:4-jet}, it is clear that if one could efficiently discriminate quark-initiated jets against gluon-initiated jets, even more stringent limits could potentially be obtained by selecting the channels with the largest $O_G$ effects. Given the recent progress in quark/gluon discrimination, see, {\it e.g.}, \cite{Gras:2017jty}, we deem this could be an interesting direction to explore in the future. 

\subsection{EFT validity for multi-jet limits}
\begin{figure}[h]
\begin{center}
\includegraphics[width=0.7\textwidth,trim=0cm 3cm 0cm 0cm]{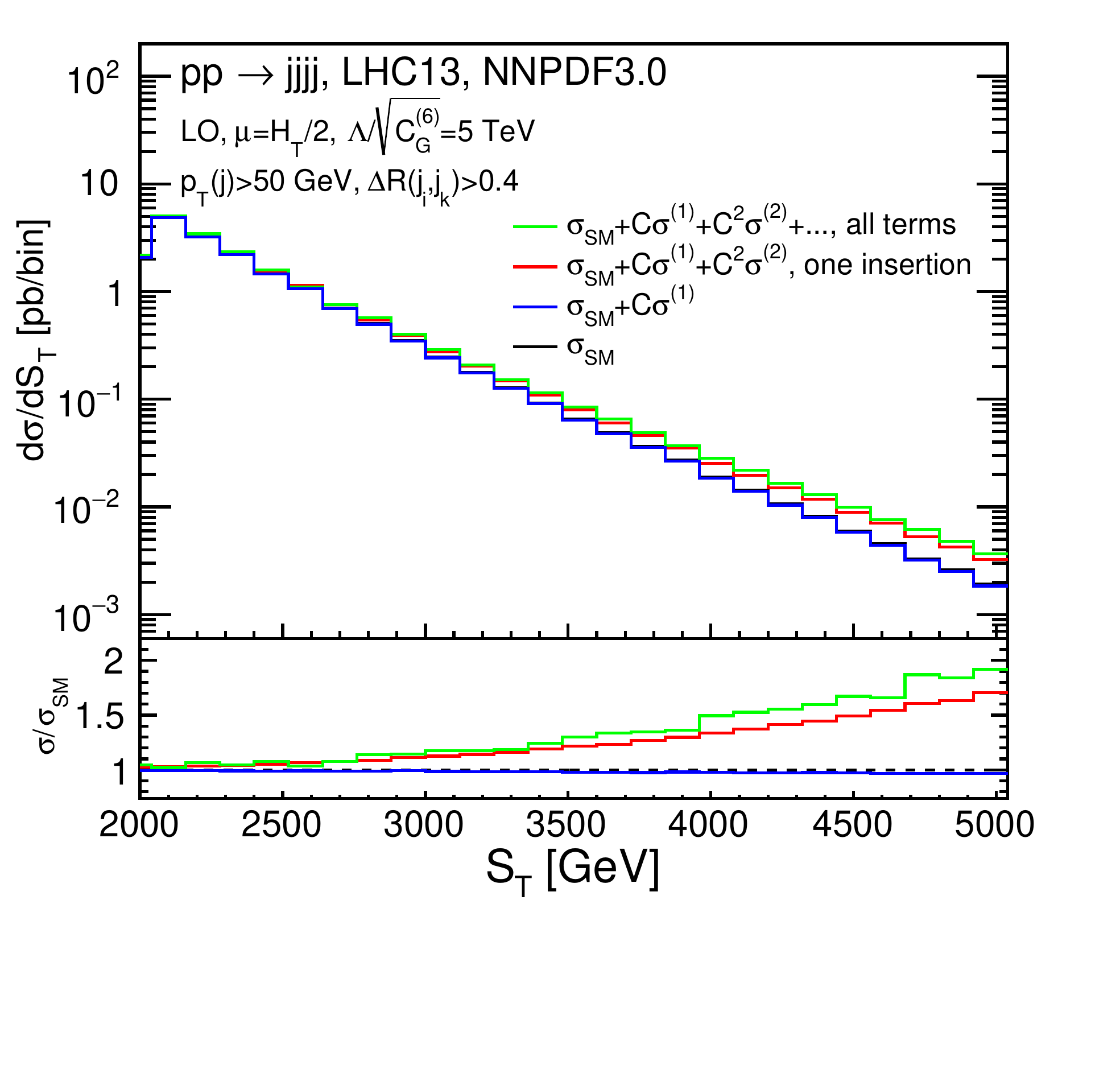}
\caption{\label{sqvsinter} Comparison between the contributions of order $\mathcal{O}(1/\Lambda^2)$ (interference), $\mathcal{O}(1/\Lambda^4)$ and higher for the production of four partonic jets at LHC13. }
\end{center}
\end{figure}

\begin{figure}[h!]
\begin{center}
\includegraphics[width=0.8\textwidth]{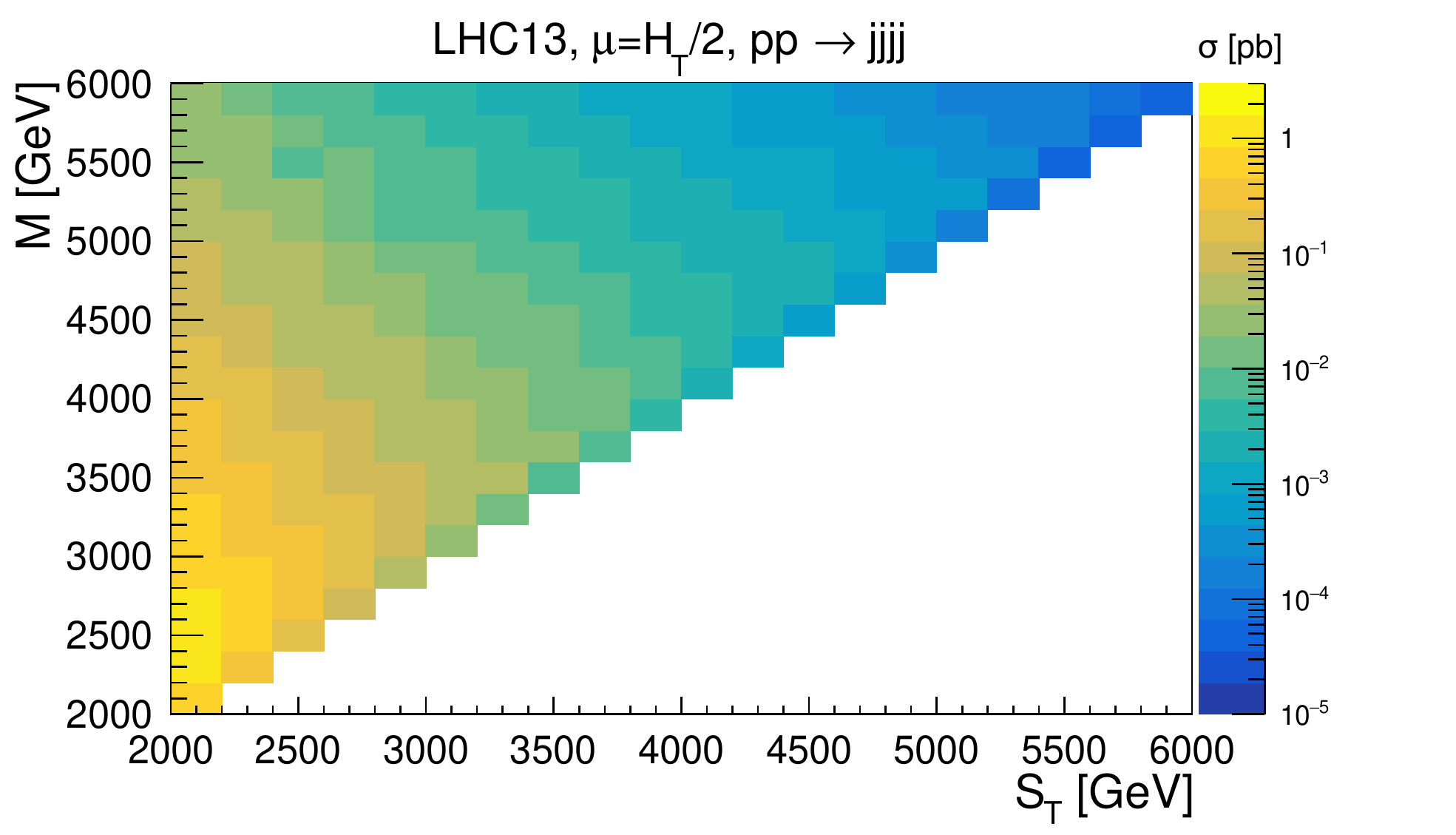}
\caption{\label{scatter}Distribution of the centre-of-mass energy (M) as a function of $S_T$ for the production of four partonic jets  at LHC13.}
\end{center}
\end{figure}

\begin{figure}[t]
\begin{center}
\includegraphics[width=0.7\textwidth,trim=0cm 3cm 0cm 0cm]{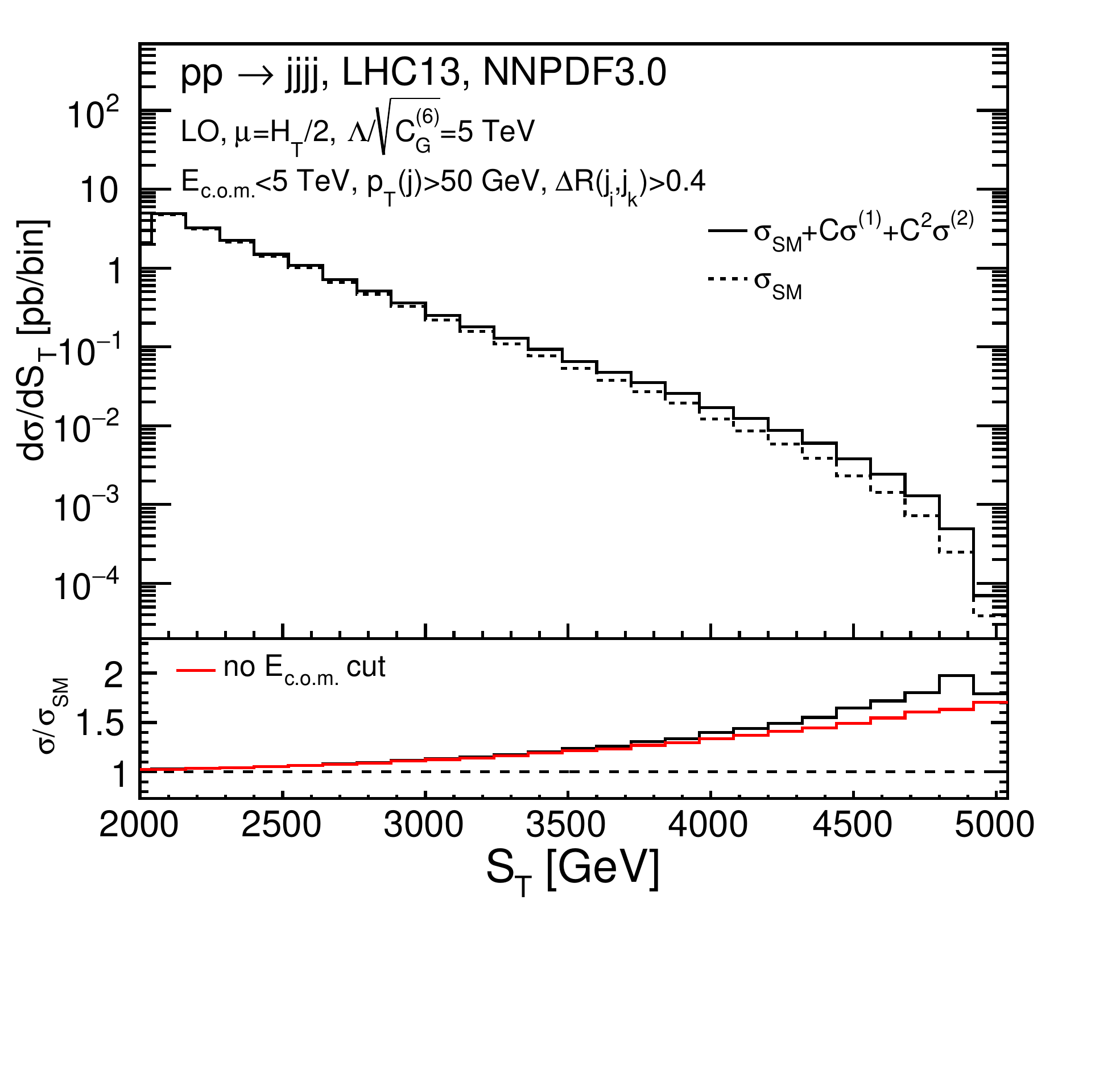}
\caption{\label{cut} Distribution of $S_T$ and corresponding deviation from the SM after applying a cut on the partonic centre of mass energy of the events, $E_{\textrm{c.o.m.}}< 5$ TeV. 
}
\end{center}
\end{figure}

The sensitivity of the $S_T$ observable on $O_G$ is driven by contributions quadratic in $1/\Lambda^2$, and possibly of even larger powers. This is demonstrated in Figure~\ref{sqvsinter} where the contributions at $\mathcal{O}(1/\Lambda^2)$, $\mathcal{O}(1/\Lambda^4)$ and above are separated. The $\mathcal{O}(1/\Lambda^4)$ contribution is growing faster with the energy and therefore deserves further attention in order to ensure that the EFT validity condition of $E<\Lambda$ remains satisfied.
The precise application of the EFT validity constraint implies insuring that the characteristic energy probed by the observable considered is smaller than $\Lambda$. This is in general a stronger constraint than just imposing $S_T<\Lambda$, since by definition we have $E_{\textrm{c.o.m.}}>S_T$ so that regions of $S_T$ below $\Lambda$ can still receive contributions from events with $E_{\textrm{c.o.m.}}>\Lambda$, as demonstrated in Figure~\ref{scatter}. 
In order to assess the impact of excluding the $E_{\textrm{c.o.m.}}>\Lambda$ region, we considered a simulation of the production of four isolated partonic jets, allowing at most one insertion of the $O_G$ operator (alike what was done in Ref.~\cite{Krauss:2016ely}) 
and using Monte Carlo truth in order to remove all events featuring $E_{\textrm{c.o.m.}}>5$ TeV. We report the results obtained with this procedure in Figure~\ref{cut}, keeping only the contributions from events passing this cut on $E_{\textrm{c.o.m.}}$. We find that the cut has no significant effect on the expected deviation w.r.t the SM and therefore conclude that the EFT expansion condition remains valid for the limit of 5 TeV set on $\Lambda$.
\subsection{Dimension-8 effects on multi-jet observables}
\begin{figure}[h!]
\begin{center}
\includegraphics[width=0.7\textwidth,trim=0cm 3cm 0cm 0cm]{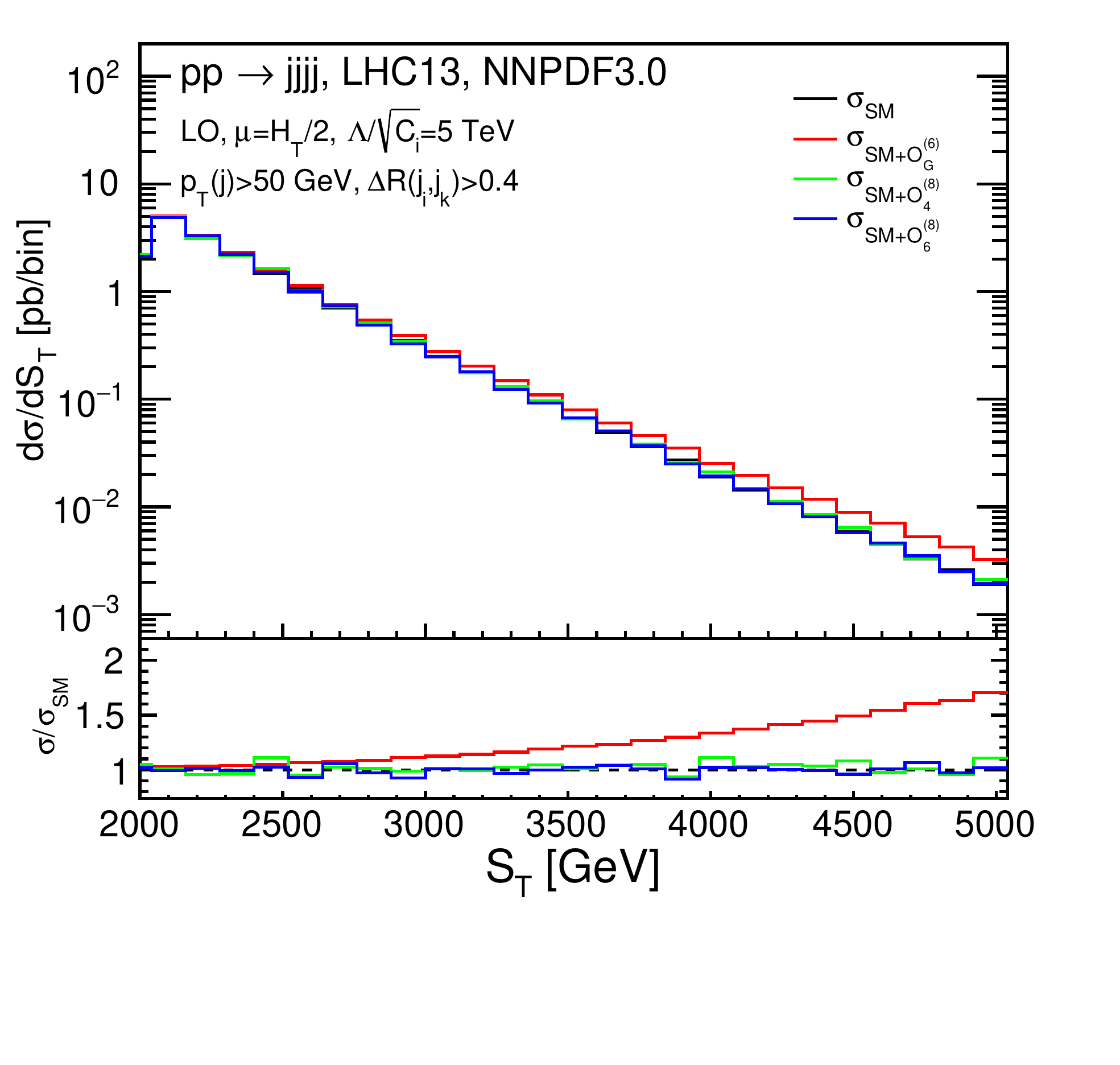}
\caption{\label{dim8}Impact of the dimension-8 operators of Eqs.~\ref{dim8operators1}-\ref{dim8operators2} on the $S_T$ observable. A maximum of one insertion of each operator is allowed at the amplitude level and all terms up to $O(1/\Lambda^8)$ are retained. }
\end{center}
\end{figure}

Although we have shown that the limit-setting analysis presented in Ref.~\cite{Krauss:2016ely} respects the EFT validity condition of probing energies smaller than $\Lambda$, we must also estimate the magnitude of the omitted contributions of order $O(1/\Lambda^4)$. Given that the contribution of $O_G$ is completely dominated by the squared terms of order $O(1/\Lambda^4)$, it is interesting to compute the contribution of the interference of dimension-8 operators with the SM amplitudes, which are formally of the same order in the $1/\Lambda^2$ expansion, in the relevant regions of phase space. To this end, we implemented the following subset of dimension-8 operators:
\begin{eqnarray}
\label{dim8operators1}
O_4^{(8)}=\frac{g_s^2}{2} G^{\mu\nu}_{a}G_{\mu\nu}^{a}G^{b}_{\rho\sigma}G^{\rho\sigma}_{b}\\
\label{dim8operators2}
O_6^{(8)}=\frac{g_s^2}{2} G^{\mu\nu}_{a}G_{\mu\nu}^{b}G^{a}_{\rho\sigma}G^{\rho\sigma}_{b},
\end{eqnarray}
and computed their contribution to the $S_T$ observable, as shown in Figure~\ref{dim8}. We find that in the $S_T$ region up to 5 TeV, the contribution from these dimension-8 operators is suppressed compared to that of $O_G$ for identical values of $\Lambda$. This completes our checks related to the validity of the EFT and confirms that the limits obtained in the original study of Ref.~\cite{Krauss:2016ely} are robust from the EFT validity standpoint. 

\subsection{Additional observables in multi-jet production}
As already argued, the variable $S_T$ measured by CMS probes mainly di-jet configurations where two jets are hard and the remaining tagged ones are soft in comparison. For these configurations, the interference term is suppressed and the squared contribution dominates. In order to investigate the contribution of the  $\mathcal{O}(1/\Lambda^2)$ interference  in configurations where \emph{all} partonic jets are hard, we consider the tree-level contribution of $O_G$ to three- and four-jet final states as a function of a minimum $p_T$-cut applied to \emph{all} partonic jets. The cross-sections of the production of three hard parton jets for different jet-$p_T$ cuts and $\Delta R=\sqrt{\Delta \phi^2 + \Delta \eta^2}=0.4$ are given in Table~\ref{tab:tj1}, for $\Lambda=5$ TeV. We observe that for all chosen values of the $p_T$-cut, the interference remains suppressed. The relative importance of this interference decreases with the $p_T$-cut. However,  this is an accidental outcome of the combination of the various partonic channels which feature a very different dependence on both the $p_T$-cut and the operator $O_G$, as already discussed in the Section~\ref{multiplicitysection}. 
%
\begin{table}[h]
\begin{center}
\renewcommand{\arraystretch}{1.25}
\begin{tabular}{c | c c c}
\hline\hline
$p_{T,{\rm min}}(j)$ & SM [pb] & $\mathcal{O}(1/\Lambda^2)$ [pb] & $\mathcal{O}(1/\Lambda^4)$ [pb]\\
\hline
50 & 8.85$\cdot 10^{+5}$ & 3.37$\cdot 10^{+1}$& 3.03$\cdot 10^{+1}$ \\
100 & 3.13$\cdot 10^{+4}$ & 5.14$\cdot 10^{0}$ & 9.99$\cdot 10^{0}$ \\
200 & 7.82$\cdot 10^{+2}$ & 5.94$\cdot 10^{-1}$ & 2.23$\cdot 10^{0}$ \\
500 & 2.44$\cdot 10^{0}$ & 1.89$\cdot 10^{-2}$ & 1.10$\cdot 10^{-1}$ \\	
1000	& 8.08$\cdot 10^{-3}$ & 4.91$\cdot 10^{-4}$ & 2.48$\cdot 10^{-3}$ \\
\hline
\end{tabular}
\caption{\label{tab:tj1} Cross-sections for the production of three partonic hard jets, isolated using different values of a $p_{T,min}$-cut and $\Delta R=0.4$. The scale $\Lambda$ is set to $5$ TeV. In the computation of the result of order $\mathcal{O}(1/\Lambda^4)$, only the contributions from at most one insertion of the operator $O_G$ in the amplitudes are considered.}
\end{center}
\end{table}
It is also interesting to investigate which helicity configurations contribute to the $gg \to ggg$ amplitude, and we find that more configurations contribute to the amplitude featuring one insertion of the $O_G$ operator compared to the pure QCD one, as was already noted in~\cite{Dixon:1993xd}. Very approximately, one should therefore expect the cross-section arising from the square of this amplitude to be dominant as it involves the summation over a larger number of positive contributions. The interference term, on the other hand, involves fewer contributions and being not positive definite its contribution drastically reduces once averaged over phase space. 

We also consider the production of four partonic jets, in the presence of the same cuts, and report our results in Table~\ref{tab:tj2}. We find that the suppression of the interference is smaller than for three partonic jets. The relative importance of the interference decreases with the partonic jet $p_T$-cut, eventually reaching $\sim10\%$ for $p_{T,min}>1000$ GeV. We conclude that, given the current constraint on the $O_G$ operator coefficient, we cannot identify any partonic jet-$p_T$ region where the interference contribution is comparable to those of order $\mathcal{O}(1/\Lambda^4)$ and significant compared to the SM ones for these semi-inclusive measurements.  

\begin{table}[h]
\begin{center}
\renewcommand{\arraystretch}{1.25}
\begin{tabular}{c | c c c}
\hline\hline
$p_{T,{\rm min}}(j)$ [GeV]  & SM [pb] & $\mathcal{O}(1/\Lambda^2)$ [pb] & $\mathcal{O}(1/\Lambda^4)$ [pb] \\
\hline
50 & 1.20$\cdot 10^{+5}$ & -2.55$\cdot 10^{+1}$ & 1.48$\cdot 10^{+1}$ \\
100 & 2.87$\cdot 10^{+3}$ & -2.45$\cdot 10^{0}$ & 2.94$\cdot 10^{0}$ \\
200 & 4.37$\cdot 10^{+1}$ & -1.55$\cdot 10^{-1}$ & 3.57$\cdot 10^{-1}$ \\
500 & 5.18$\cdot 10^{-2}$ & -1.04$\cdot 10^{-3}$ & 5.85$\cdot 10^{-3}$ \\
1000	& 4.55$\cdot 10^{-5}$ & -3.89$\cdot 10^{-6}$ & 3.60$\cdot 10^{-5}$ \\
\hline
\end{tabular}
\caption{\label{tab:tj2} Cross-sections for the production of four partonic hard jets, isolated using different values of a $p_{T,min}$-cut and $\Delta R=0.4$. The scale $\Lambda$ is set to $5$ TeV. In the computation of the result of order $\mathcal{O}(1/\Lambda^4)$, only the contributions from at most one insertion of the operator $O_G$ in the amplitudes are considered.}
\end{center}
\end{table}


\begin{figure}[h!]
\begin{center}
\includegraphics[width=0.6\textwidth]{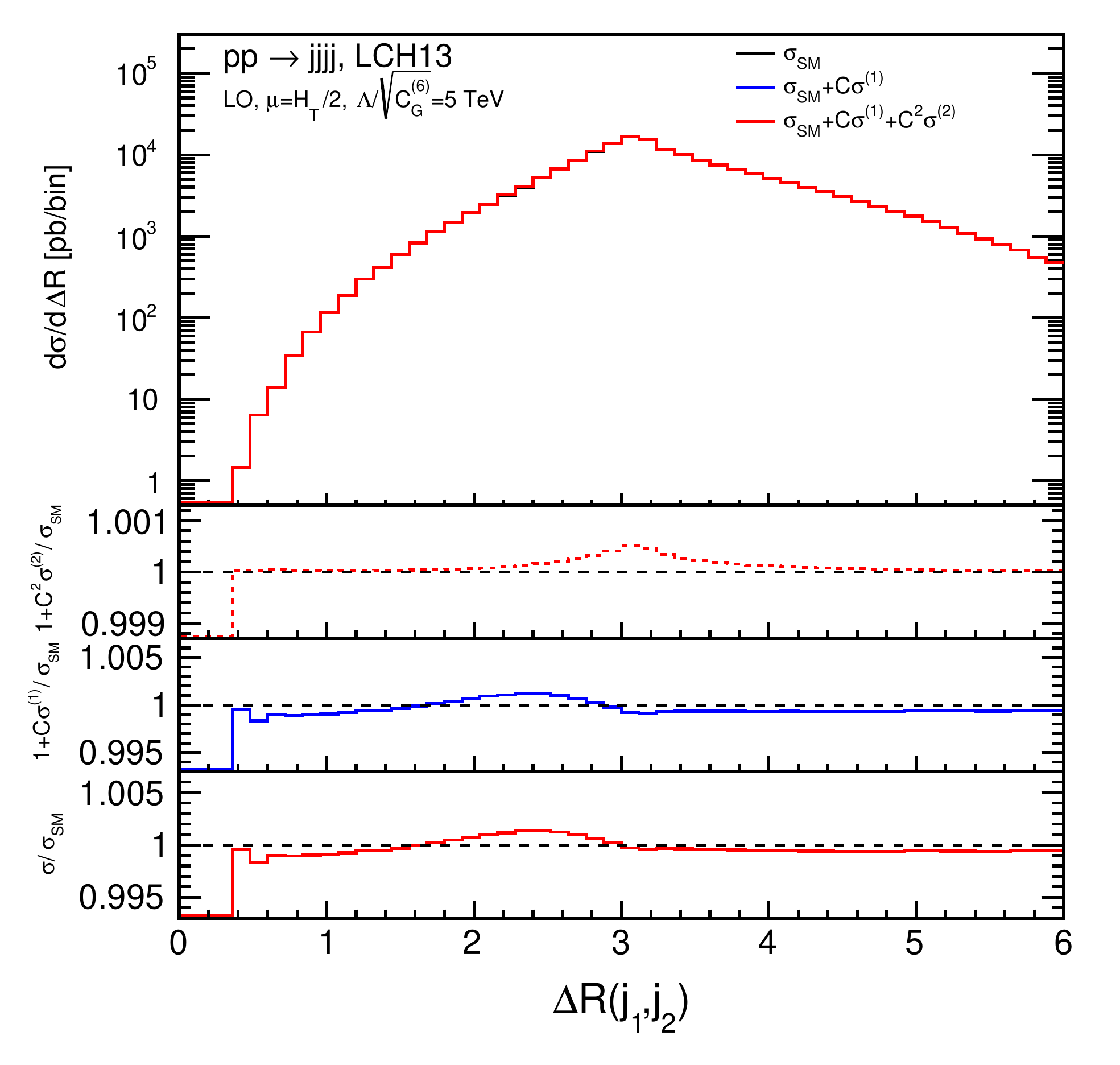}
\caption{\label{drjj}
Angular separation $\Delta R = \sqrt{ \Delta \phi^2 + \Delta \eta^2}$ between the two leading partonic jets in 4-jet production, for the SM and the signal of order $\mathcal{O}(1/\Lambda^2)$ and $\mathcal{O}(1/\Lambda^4)$, in the presence of the cuts $S_{T}$>2 TeV, $\Delta R>0.4$ and $p_{T,min}>50$ GeV (applied to all jets). In the computation of the result of order $\mathcal{O}(1/\Lambda^4)$, only the contributions from at most one insertion of the operator $O_G$ in the amplitudes are considered.}
\end{center}
\end{figure}
\begin{figure}[h!]
\begin{center}
\includegraphics[width=0.6\textwidth]{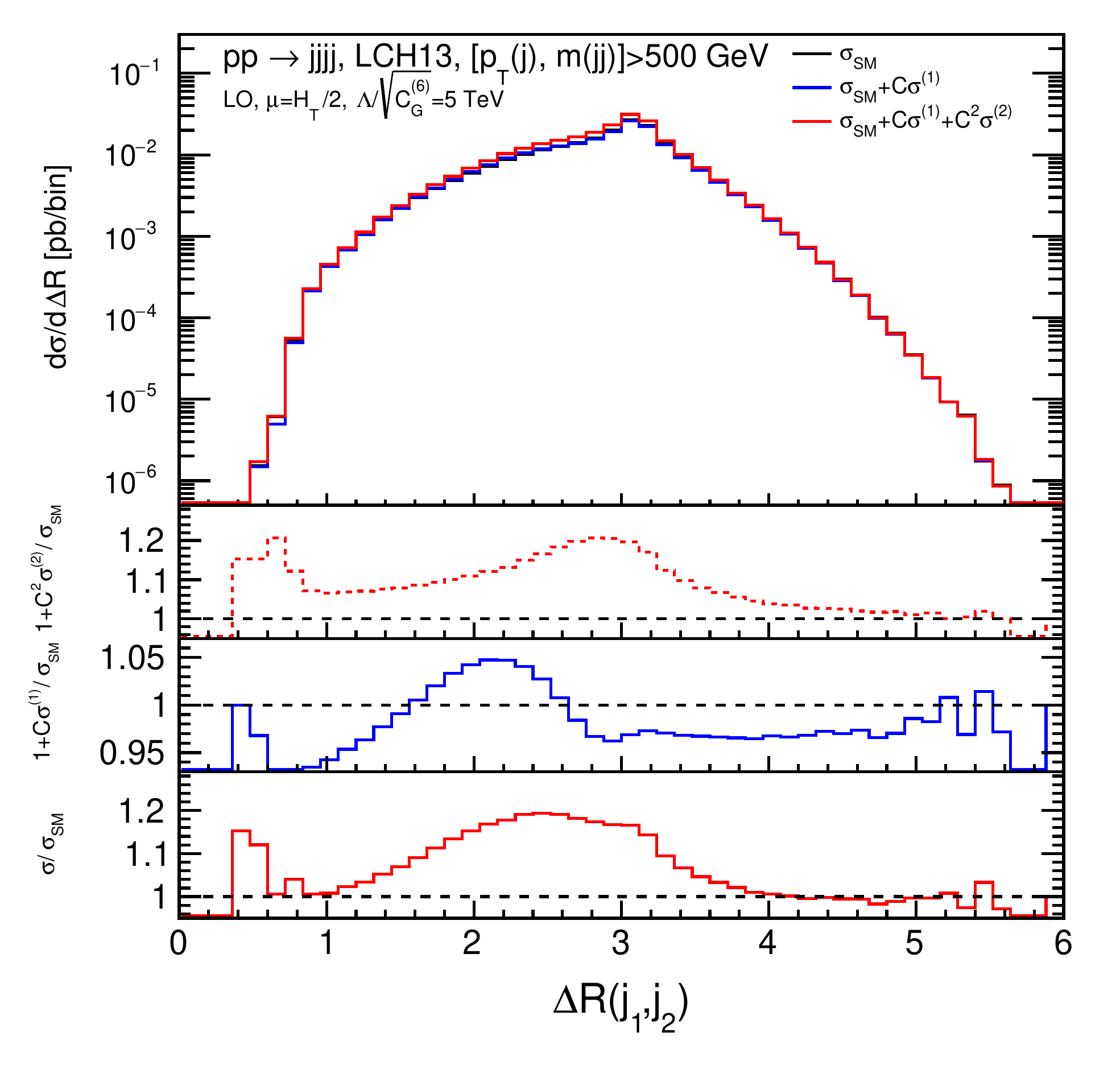}
\caption{\label{drjjcut} Angular separation $\Delta R = \sqrt{ \Delta \phi^2 + \Delta \eta^2}$ between the two leading partonic jets in 4-jet production, for the SM and the signal of order $\mathcal{O}(1/\Lambda^2)$ and $\mathcal{O}(1/\Lambda^4)$, in the presence of the cuts $p_{T,min}> 500$ GeV and $M_{jj}>500$ GeV.}
\end{center}
\end{figure}

We note that for some observables the interference changes sign, for example in the angular separation $\Delta R$ between the two leading jets in the production of four partonic jets, shown in Figure~\ref{drjj}. In this observable, the interference dominates over the squared contribution but the deviation from the SM is below the percent level. The overall contribution of the $O_G$ operator can be enhanced when considering additional cuts ($p_{T,min}> 500$ GeV, $M_{jj}>500$ GeV) restricting the kinematic configurations to four hard jets, as shown in Figure~\ref{drjjcut}. Although the signal can reach 20\%  of the very small fiducial SM yield  (for $2 < \Delta R < 3$), it is again dominated by contributions of order $\mathcal{O}(1/\Lambda^4)$.

In light of the above results, one is lead to the conclusion that  only when considering relatively simple observables such as the transverse momenta of jets or their angular separation, we always find ourselves in a situation where the signal induced by the $O_G$ EFT operator is either too small to ever be used for placing bounds on $\Lambda$ at LHC13, or it is dominated by contributions of order $\mathcal{O}(1/\Lambda^4)$ which are potentially more sensitive to a breakdown of the EFT validity.

\section{Impact of the multi-jet bound on $O_G$ in other observables}
Having established the validity of the constraints set on $O_G$ from high-multiplicity jet events, in this section we re-analyse the processes previously suggested in the literature to constrain $O_G$. We consider in turn heavy quark production and higher order QCD corrections to di-jet production. In particular, we compute the 1-loop corrections to the interference between the SM and $O_G$ di-jet amplitudes and study the angular observables in 3-jet events as suggested in \cite{Dixon:1993xd}.
\subsection{Constraining $O_G$ in heavy quark production}
\begin{figure}[h!]
\begin{center}
\includegraphics[width=0.6\textwidth,trim=0cm 0cm 0cm 0cm]{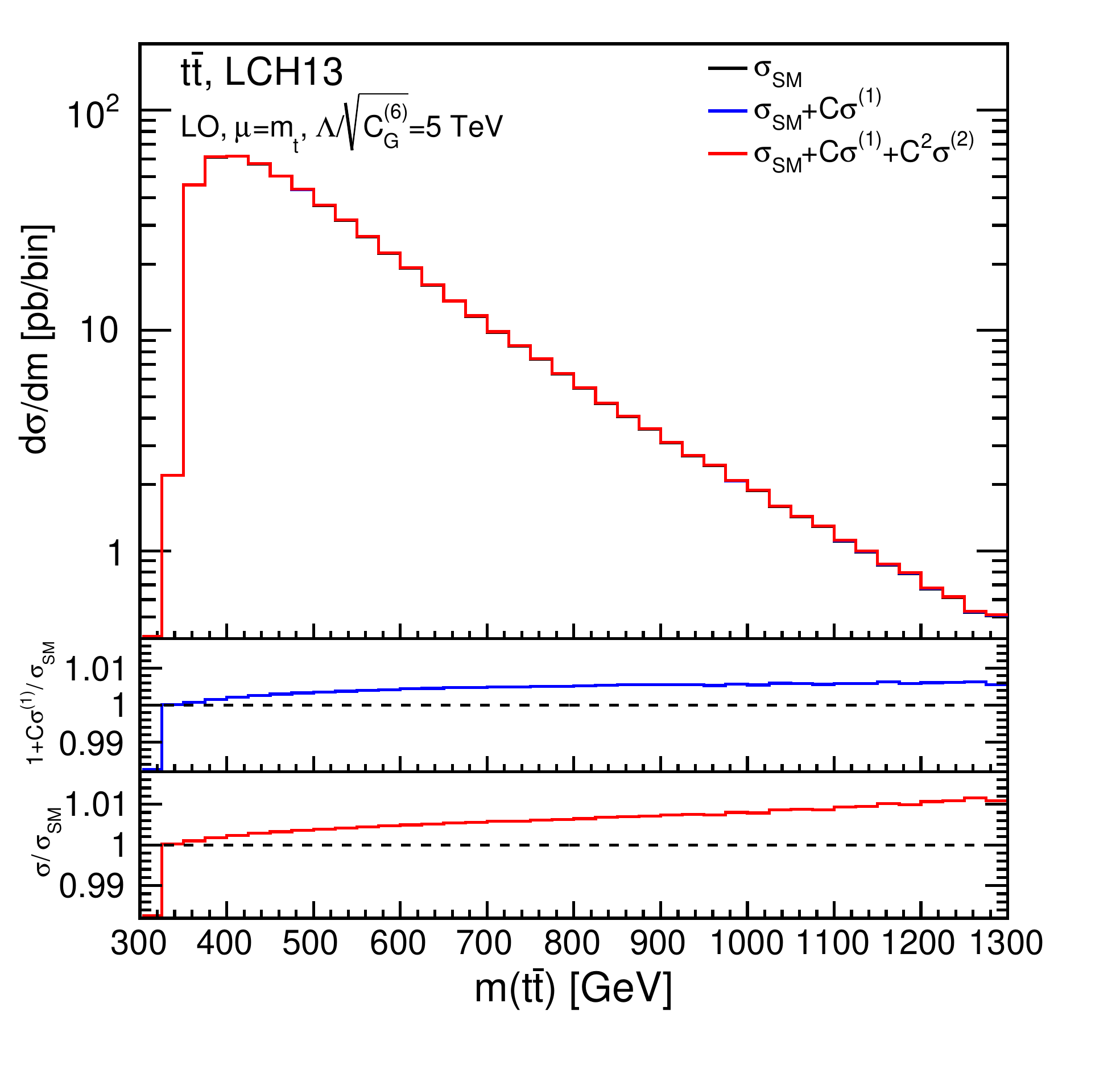}
\caption{\label{ttx} Invariant mass of a pair of top quarks produced at LHC13, for the SM and including at most one insertion of the $O_G$ operator  in the production amplitudes ($\mathcal{O}(1/\Lambda^2)$ and $\mathcal{O}(1/\Lambda^4)$ terms).  }
\end{center}
\end{figure}
\begin{figure}[h!]
\begin{center}
\includegraphics[width=0.6\textwidth,trim=0cm 0cm 0cm 0cm]{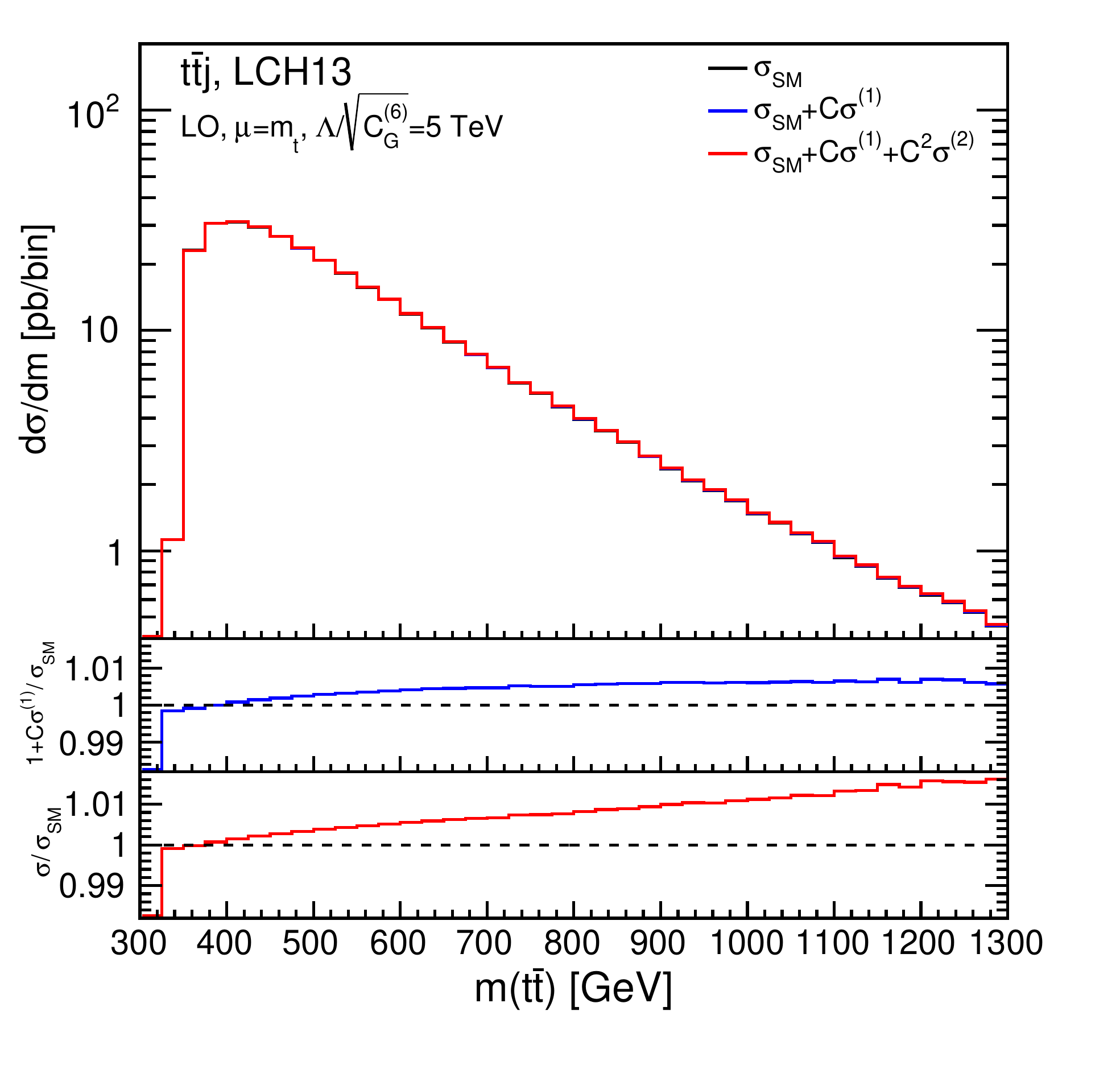}
\caption{\label{ttxj} Invariant mass of a pair of top quarks produced in association with a jet at LHC13, for the SM and including at most one insertion of the $O_G$ operator  in the production amplitudes ($\mathcal{O}(1/\Lambda^2)$ and $\mathcal{O}(1/\Lambda^4)$ terms). A transverse momentum cut of $p_{T,min}>50$ GeV is applied to the jet produced. }
\end{center}
\end{figure}
Massive quark pair-production opens up additional non-zero helicity configurations common between the amplitudes with and without one insertion of the $O_G$ operator~\cite{Cho:1993eu,Cho:1994yu}. The interference contribution of order $\mathcal{O}(\Lambda^2)$ is then once again resurrected and we consider its possible phenomenological applications in this section. Its impact in the context of top-quark pair production was first studied in ref.~\cite{Cho:1993eu,Cho:1994yu}. We re-examine the situation here in light of the constraints found in ref.~\cite{Krauss:2016ely} and considering the production of a pair of top quarks, also in association with one QCD jet. We present our results in Figures \ref{ttx}-\ref{ttxj} and observe that the interference contribution of order $\mathcal{O}(\Lambda^2)$ is indeed not suppressed compared to that of $\mathcal{O}(\Lambda^4)$, but its overall signal strength normalised to the SM yield is at the percent level in the tail of the distributions. This implies that given the current experimental and theoretical control over differential measurements in top quark pair production, this process cannot improve on the current limit of $\Lambda=5$ TeV for the $O_G$ operator. 

\begin{figure}[h!]
\begin{center}
\includegraphics[width=0.7\textwidth,trim=0cm 2cm 0cm 0cm]{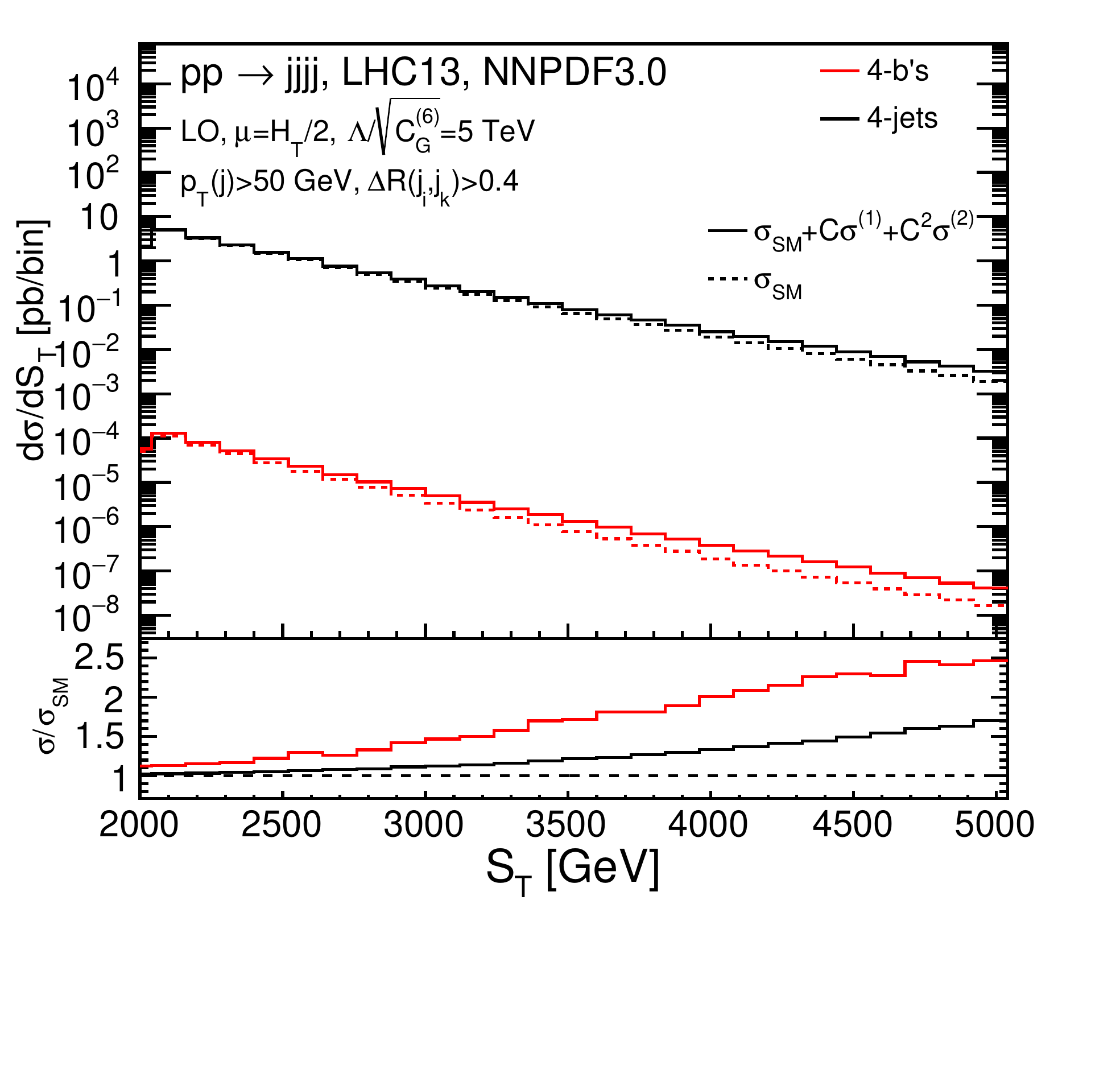}
\caption{\label{4b} Comparison of the $S_T=\cancel{E_T}+\sum^{N_{\textrm{jets}}}_{j=1} E_{T,j}$ (see Eq.~\ref{STdef}) distribution in four b-quark production (red) and four-jet production (black). }
\end{center}
\end{figure}

Finally, we also consider the production of four $b$-quarks, as shown in Figure~\ref{4b}, revisiting the same observable $S_T$ discussed in sect.~\ref{multiplicitysection} and in the high energy region. The selection of this process is motivated by the fact that we found the $O_G$ contribution to be enhanced in the multi-jet production channel $gg\to qqqq$ (see Table \ref{table:4-jet}). We find that the presence of the $b$-quark mass indeed renders the $O_G$ contribution larger for four $b$-jet production than inclusive four-jets, but the corresponding production rate is too small. Indeed, the suppression of more than four orders of magnitude w.r.t to four-jet production, in combination with the eventual need of accounting for $b$-quark tagging efficiencies, prevents this channel from providing significant constraints on the triple gluon operator.

\subsection{One-loop corrections to the $O_G$ contribution to di-jet production}
\label{oneloopstudy}
Up to this point we have established that the contribution of the interference of order $\mathcal{O}(1/\Lambda^2)$ is zero for tree-level di-jet production and suppressed compared to those of order $\mathcal{O}(1/\Lambda^4)$ for a variety of multi-jet observables. In this section we investigate whether the 1-loop corrections of order $\mathcal{O}(1/\Lambda^2)$ to di-jet production can lift the suppression of the interference.

The helicity structure of the $gg\rightarrow gg$ and $gg \rightarrow q\bar{q}$ amplitudes with exactly one $O_G$ insertion are orthogonal to the pure QCD ones. This implies that the interference term is exactly zero at the tree level (see Appendix~\ref{analyticGGG} for details). One-loop amplitudes for these processes, both with and without a single insertion of the $O_G$ operator, open up additional helicity configurations yielding non-zero interfering contributions.
Similarly, considering one more parton in the final state turns on new interfering helicity configurations. These two contributions can be thought of as the usual virtual and real-emission pieces of the NLO QCD corrections to di-jet production. We however refrain from employing this terminology since the tree-level Born contribution of order  $\mathcal{O}(1/\Lambda^2)$ vanishes. An important consequence of this is that, by virtue of the factorisation properties of UV and IR divergences, both these inclusive `virtual' and `real-emission' contributions are \emph{separately} UV and IR finite\footnote{We stress that the amplitudes for $p p \rightarrow j j j$ at order $\mathcal{O}(1/\Lambda^2)$ still feature \emph{integrable} IR singularities, implying that local counterterms are still necessary in order to numerically perform an inclusive computation.}
making the former akin to a loop-induced computation.

Using {\sc FeynRules}~\cite{Alloul:2013bka} together with the {\sc NLOCT}~\cite{Degrande:2014vpa} module, we built a {\sc UFO}~\cite{Degrande:2011ua} model\footnote{The model is made publicly available through the online model repository of {\sc MG5aMC} v2.6.1+, under the name {\tt GGG\_EFT\_up\_to\_4point\_loops}, and at \url{http://feynrules.irmp.ucl.ac.be/wiki/NLOModels}. We stress that this model is not suited for computations of $n$-points EFT loops, with $n>4$.} containing the necessary $R_2$ counterterms\footnote{$R_2$ counterterms reproduce the rational part of one-loop amplitudes that originate from the $d-4$ dimensional part of the loop numerator and which can therefore not be reproduced by purely numerical codes working in four dimensions.} allowing {\sc MadLoop}~\cite{Hirschi:2011pa,Alwall:2014hca} to compute the one-loop di-jet contribution of order $\mathcal{O}(1/\Lambda^2)$. As the finite one-loop amplitudes for di-jet production with exactly one insertion of the $O_G$ operator are computed here for the first time, we provide more details on some analytical results obtained in the case of the four-gluon one-loop amplitude in appendix~\ref{analyticGGG}.

We now turn to discussing the phenomenological relevance of the finite one-loop di-jet contributions of order $\mathcal{O}(1/\Lambda^2)$, computed within the loop-induced module~\cite{Hirschi:2015iia} of the {\sc MG5\_aMC}~\cite{Alwall:2014hca} framework. Numerical results for this contribution and for different cuts on partonic jets $p_T$ can be found in Table~\ref{tab:nlodi-jet} in Appendix \ref{app:oneloop}.

\begin{figure}[h!]
\begin{center}
\includegraphics[height=8cm]{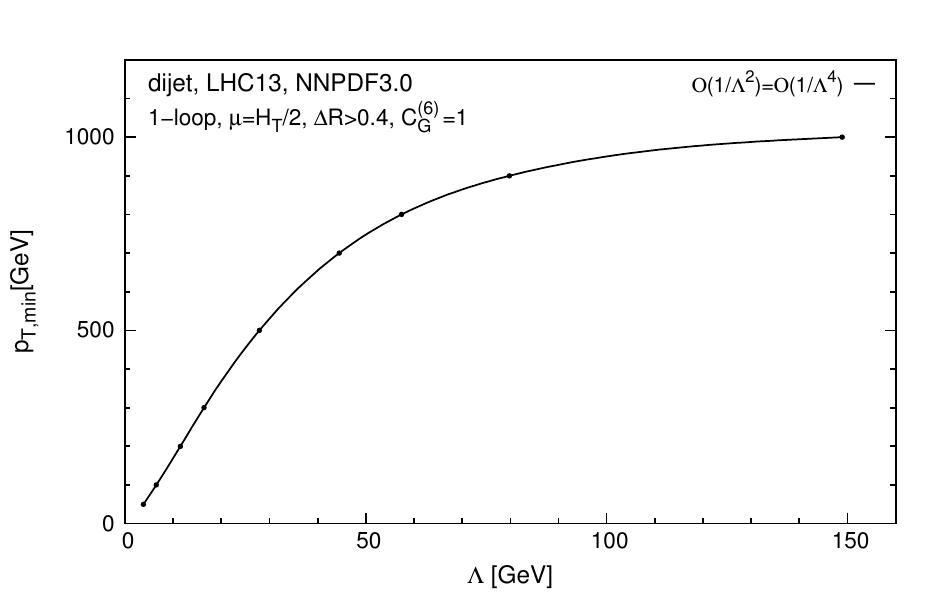}
\caption{\label{ptatequal}Value of the partonic jet cut $p_{T,min}(j)$ at which the contribution of order $\mathcal{O}(1/\Lambda^4)$ exceeds the one of the one-loop interference term of order $\mathcal{O}(1/\Lambda^2)$.}
\end{center}
\end{figure}

We find quite large cancellations between the $\mathcal{O}(1/\Lambda^2)$ contributions of the various partonic subprocesses and the $\mathcal{O}(1/\Lambda^4)$ term happens to dominate over the interference for all $p_T$ regions shown for $C_G=1$ and $\Lambda=5$ TeV.
Given their different scaling with the characteristic energy probed, it is interesting to study the value of the cut $p_{T,min}$ at which the contributions of order $\mathcal{O}(1/\Lambda^2)$ and $\mathcal{O}(1/\Lambda^4)$ are equal to each other for different values of the energy scale $\Lambda$. We report our results in Figure~\ref{ptatequal} which shows that even for $\Lambda$ as large as $150$ TeV, the squared contribution $\mathcal{O}(1/\Lambda^4)$ dominates when restricting the jet kinematics to $p_T(j)>1$ TeV.\footnote{Whilst we show results as a function of $p_{T,min}$, similar conclusions can be drawn if we consider slices of $p_T$, as the jet $p_T$ spectrum is steeply falling.}


\begin{figure}[h!]
\begin{center}
\includegraphics[height=8cm]{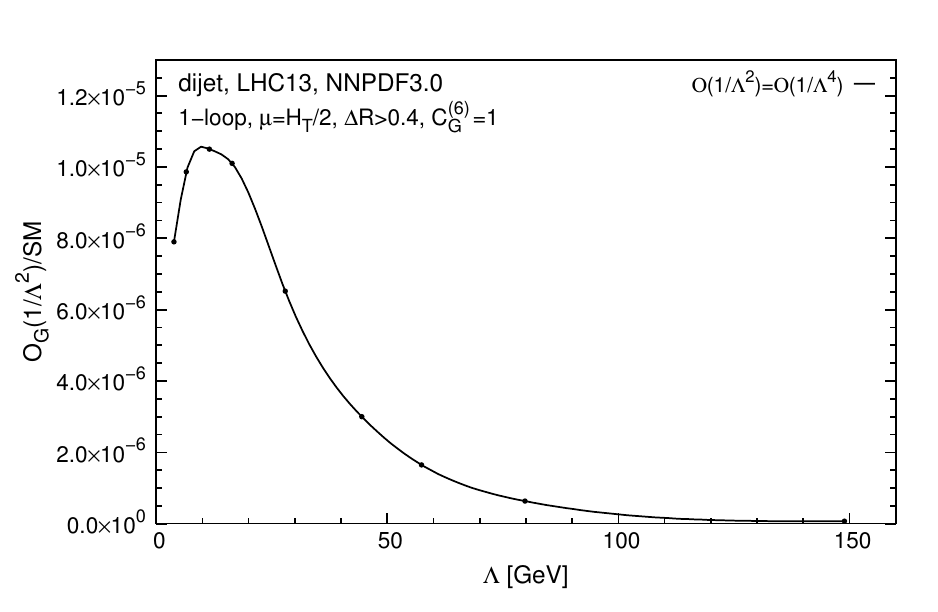}
\caption{\label{limits} Contributions from the one-loop interference contribution of order $\mathcal{O}(1/\Lambda^2)$, normalised to the SM yield and in presence of a minimum partonic jet cut $p_{T,min}$ set at the value for which the two contributions of order $\mathcal{O}(1/\Lambda^2)$ and $\mathcal{O}(1/\Lambda^4)$ are equal to each other (see Figure~\ref{ptatequal}).}
\end{center}
\end{figure}

We stress that signal strength alone is not meaningful, as it must be normalised to the background yield from which it ultimately needs to be separated. In the low $p_T$ region where contributions of order $\mathcal{O}(1/\Lambda^2)$ dominate, better experimental statistics may allow for somewhat weaker signal strength. We  investigate this and conclude that this is unfortunately not the case by showing in Figure~\ref{limits}, for various values of $\Lambda$, the $\mathcal{O}(1/\Lambda^2)$ contribution normalised to the SM yield and in presence of a minimum partonic jet cut  $p_{T,min}$ set at the value for which the two contributions of order $\mathcal{O}(1/\Lambda^2)$ and $\mathcal{O}(1/\Lambda^4)$ are equal to each other (see Figure~\ref{ptatequal}).
We find that even in the best case scenario ($\Lambda \sim 15$ TeV), the interfering contribution of order $\mathcal{O}(1/\Lambda^2)$ is $10^{-5}$ smaller than the SM yield, making an accurate measurement extremely challenging.

From the above, we are lead to conclude that the one-loop interference contribution of order $\mathcal{O}(1/\Lambda^2)$ in di-jet production has no phenomenological relevance at LHC13, as it is always trumped by the squared contribution of order $\mathcal{O}(1/\Lambda^4)$ in phase-space regions featuring a strong enough signal.

 \subsection{Angular observables in three-jet production}
In this section we investigate in more detail angular observables that can enhance the contribution of order $\mathcal{O}(1/\Lambda^2)$ to three-jet production at LHC13. Three-jet production can be viewed as the \emph{inclusively finite} `real-emission' counterpart of the one-loop interference `virtual' contribution to di-jet discussed in section~\ref{oneloopstudy}. The authors of ref.~\cite{Dixon:1993xd} noticed and discussed in great detail the peculiar collinear limits of the $\mathcal{O}(1/\Lambda^2)$ interference tree-level contribution to three-jet production. More specifically, they proposed observables taking advantage of this non-trivial behaviour under azimuthal rotations of two almost collinear jets to discriminate this interference contribution against the QCD background. Their study was intended for the Tevatron, and in this section we revisit their proposal in the context of the LHC13 so as to determine if such an analysis has the potential of providing additional constraints on the $O_G$ operator.

The observations of \cite{Dixon:1993xd} are based on the angle $\varphi$, the angle associated with azimuthal rotation of two collinear momenta around the direction of their sum in the c.o.m. frame. Contrary to the pure QCD matrix elements, those of the $\mathcal{O}(1/\Lambda^2)$ interference feature a non-trivial $\varphi$-dependence which is correlated with the polarisation vector of the splitting parton. The singular part of the three-jet amplitude vanishes upon integrating over $\varphi$, rendering the \emph{inclusive} three-jet cross-section of order $\mathcal{O}(1/\Lambda^2)$ free of soft and collinear divergences. The particular observable suggested in ref.~\cite{Dixon:1993xd} is the expectation value of $\langle\cos(2\varphi)\rangle$ defined as:
\begin{equation}
\langle \cos (2 \varphi) \rangle = \dfrac{\sum_i w_i \, \cos (2 \varphi )}{\sum_i w_i},
\end{equation}
where $\sum_i$ denotes the sum over all events, each with weight $w_i$. Ref.~\cite{Dixon:1993xd} also studied the distributions of the energy fraction of the most energetic jet $x_3 = \dfrac{2 E_3}{\sqrt{\hat s}}$, the angle $\psi$ between the jets plane and the beam, and the centre of mass energy for well separated jets, all of which are very mildly affected by $O_G$. They concluded that probing the collinear region was key for finding enhancement of the $O_G$ contribution, and showed that $\langle\cos(2\varphi)\rangle$ could indeed efficiently discriminate the signal against the QCD background.
 
However, this highly collinear region is hard to access experimentally as it requires the identification of two almost collinear jets. It is therefore convenient to open up the region of phase-space probed so as to make it experimentally more accessible, while at the same time retaining the discriminating power obtained from the sensitivity to the azimuthal angle $\varphi$.
This is what is achieved by the set of cuts labelled in `C' in ref.~\cite{Dixon:1993xd} and repeated here:
\begin{itemize}
\item $\sqrt{\hat s} \ge 250$ GeV
\item $x_3 \leq 0.95$ on the leading jet and $x_5 \ge 0.3$ on the trailing jet $\hfill \refstepcounter{equation}(\theequation)\label{setCcuts}$
\item $| \cos\theta | \leq 0.8$, $30^{\circ} \leq \Psi \leq 150^{\circ}$ and $x_4 \sin \theta_{34} \ge \dfrac{5 \text{ GeV}}{\sqrt{\hat s}_{\text{min}}/2}$,

\end{itemize}
where the outgoing jets (3,4,5) are ordered using the energy fractions $x_i=\dfrac{2E_i}{\sqrt{\hat s}}, i=3,4,5$ in the c.o.m. frame of the three jets, $\theta$ is the angle between the leading jet and the beam axis, $\Psi(=\varphi+\pi/2)$ the angle between the plane defined by the leading jet direction and the beam direction and the plane of the two sub-leading jets and finally $\theta_{34}$ the angle between the leading and sub-leading jet. The cuts guarantee that the transverse momenta of the two sub-leading jets are higher than 5 GeV. With the cut on $x_3$, we avoid the extremely soft and collinear region for the two sub-leading jets and the cut on $\Psi$ avoids collinearity with the beam direction. 

\begin{figure}[t]
\begin{center}
\includegraphics[width=0.6\textwidth,trim=0cm 0cm 0cm 0cm]{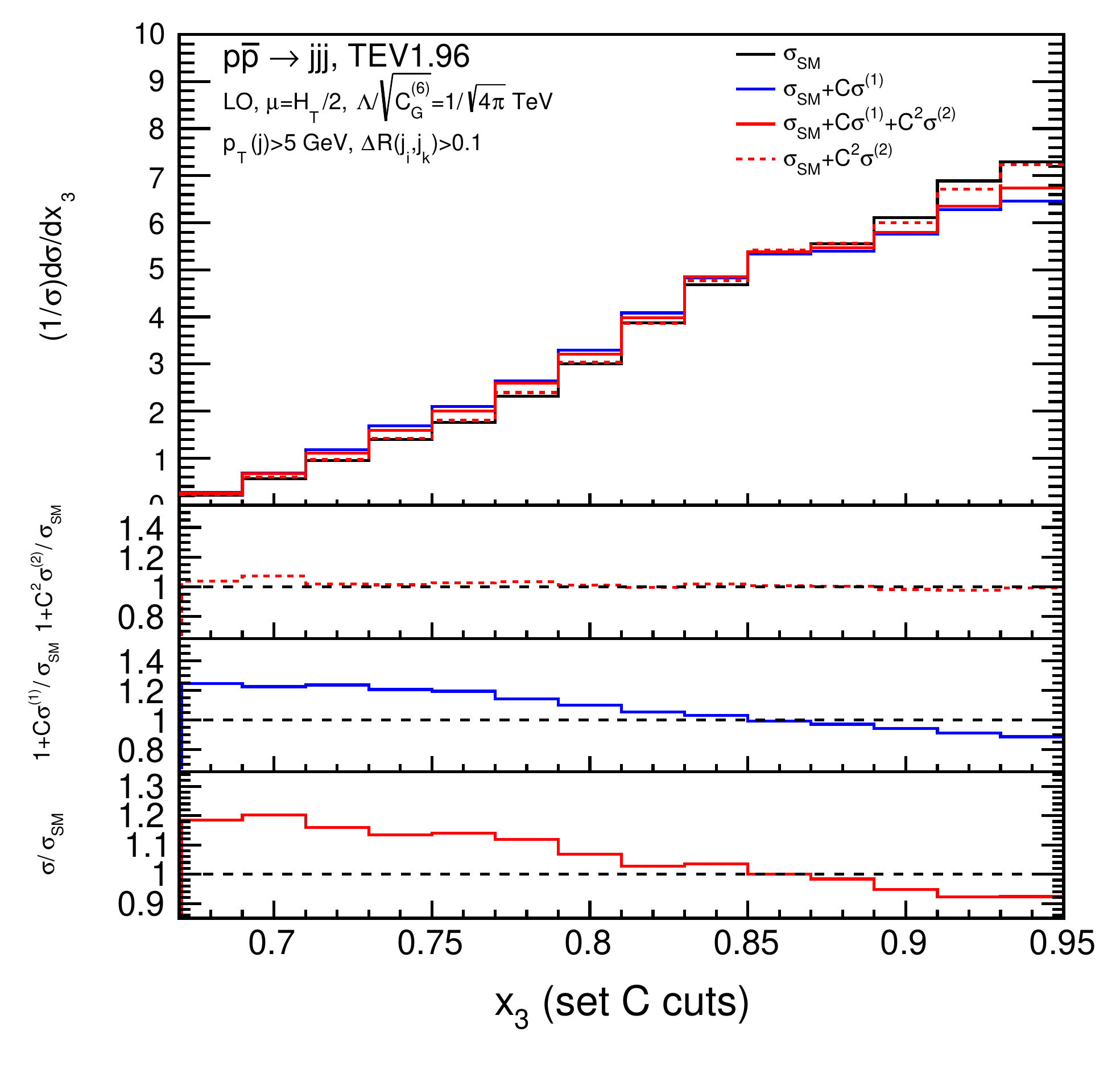}
\includegraphics[width=0.6\textwidth,trim=0cm 0cm 0cm 0cm]{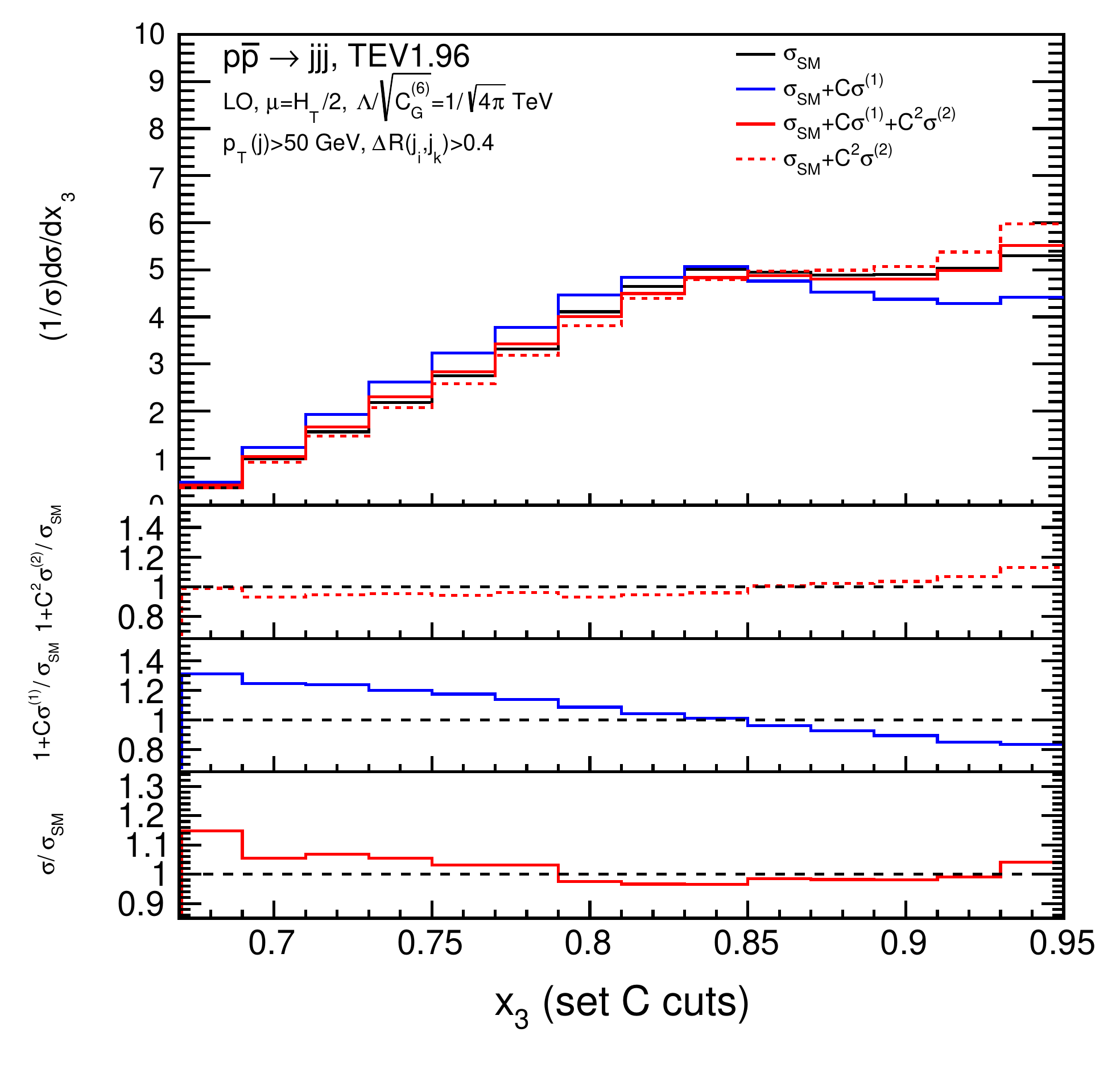}
\caption{\label{figure4Dixon} Distribution of $x_3$ in three-jet production with the cuts defined in Eq.~\ref{setCcuts}. Additional cuts of $p_T(j)>5$ GeV and $\Delta R>0.1$ are applied in the case of the upper plot, while for the lower plot the more realistic cuts of $p_T(j)>50$ GeV and $\Delta R>0.4$ are considered. }
\end{center}
\end{figure}

Here we reproduce the results of \cite{Dixon:1993xd} and examine whether the promising results found in ref.~\cite{Dixon:1993xd} remain relevant at the LHC with more realistic cuts and also given the bounds set by \cite{Krauss:2016ely}.  The behaviour of the SM and interference contributions as a function of $x_3$ for the Tevatron (for 1.96 TeV instead of 1.8 TeV used in ref.~\cite{Dixon:1993xd}) is shown in Figure \ref{figure4Dixon} (top) for the values of $C_G$ and $\Lambda$ used originally in ref.~\cite{Dixon:1993xd}. In this plot we also include the contribution of the operator at $\mathcal{O}(1/\Lambda^4)$ which was not computed in the original study.

As a first step towards a more realistic analysis, we consider additional \emph{modern} cuts, insuring that the phase-space region considered is at least experimentally accessible, i.e.:
\begin{itemize}
\item $p_T^j> 50$ GeV and $\Delta R>0.4$. $\hfill \refstepcounter{equation}(\theequation)\label{setCcutsadd}$
\end{itemize}These further cuts reduce the number of events in the collinear region as shown in Figure \ref{figure4Dixon} (bottom), again for the Tevatron. Whilst the shape of the distribution changes with the additional cuts, we find that the ratios over the SM prediction are not significantly altered. 

\begin{figure}[t]
\begin{center}
\includegraphics[width=0.6\textwidth,trim=0cm 0cm 0cm 0cm]{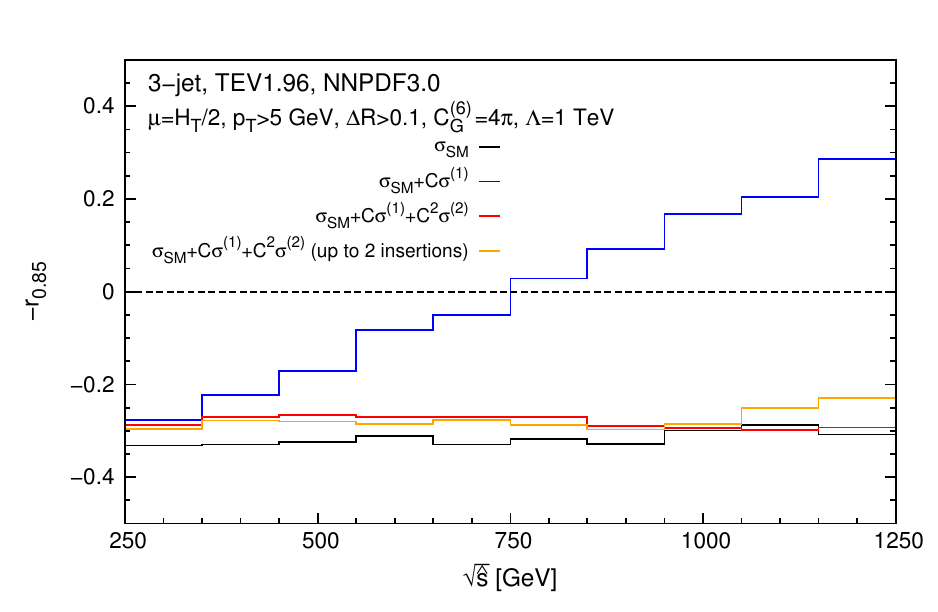}
\includegraphics[width=0.6\textwidth,trim=0cm 0cm 0cm 0cm]{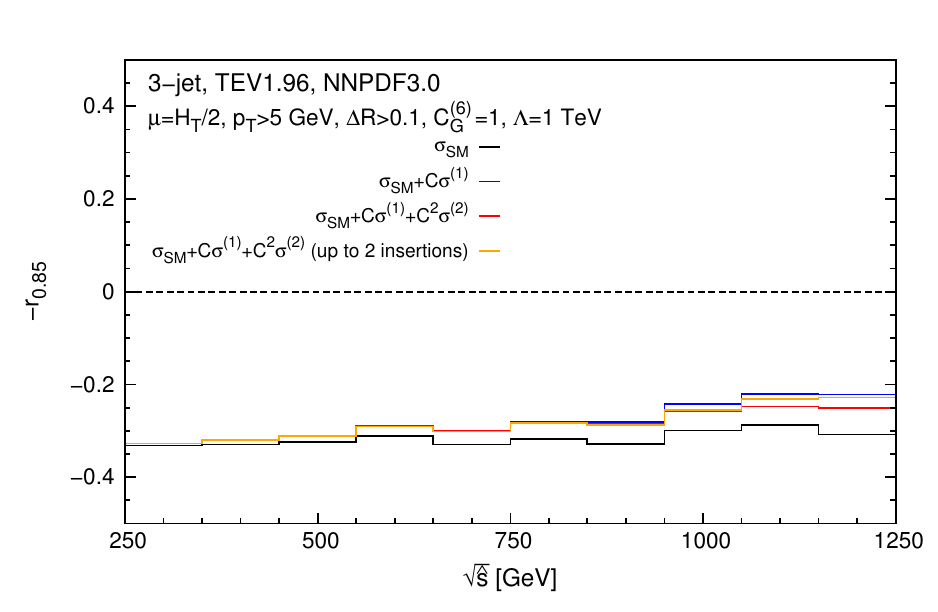}
\caption{\label{figure5Dixon} Distribution of $r_{0.85}$ as defined in Eq.~\ref{rdefDixon}, as a function of the centre-of-mass energy in three-jet production with the cuts defined in Eq.~\ref{setCcuts}, and for $C_G=4 \pi$ (upper plot) and $C_G=1$ (lower plot).  }
\end{center}
\end{figure}

In Figure \ref{figure4Dixon}, we note that the $O_G$ interference changes sign as we approach the collinear region. In ref.~\cite{Dixon:1993xd}, this feature was taken advantage of by considering the ratio $r_{0.85}$\footnote{We note here that to reproduce the results of~\cite{Dixon:1993xd} an overall minus sign had to be added to Eq.~\ref{rdefDixon} relative to the definition used in the paper. We believe this minus sign to be a typo in the original paper.}:
\begin{equation}
r_{0.85}=\dfrac{\left(\dfrac{d\sigma}{d\sqrt{\hat{s}}}\right)_{0.85<x_3<0.95}-\left(\dfrac{d\sigma}{d\sqrt{\hat{s}}}\right)_{0.75<x_3<0.95}}{\left(\dfrac{d\sigma}{d\sqrt{\hat{s}}}\right)_{0.85<x_3<0.95}+\left(\dfrac{d\sigma}{d\sqrt{\hat{s}}}\right)_{0.75<x_3<0.95}}.
\label{rdefDixon}
\end{equation}
As a validation exercise, we reproduce in Figure~\ref{figure5Dixon} the results of ref.~\cite{Dixon:1993xd} for the differential distribution of $r_{0.85}$ as a function of the c.o.m energy at the Tevatron at 1.96 TeV, for $C_G=1$  and $C_G=4\pi$ (the latter being the choice made in ref.~\cite{Dixon:1993xd}). In doing this, we also include the  $\mathcal{O}(1/\Lambda^4)$ terms, which turns out to almost completely cancel the linear contribution from the $O_G$ operator on this observable and for the original value of the coefficient used in~\cite{Dixon:1993xd}. 

\begin{figure}[h!]
\begin{center}
\includegraphics[width=0.6\textwidth,trim=0cm 0cm 0cm 0cm]{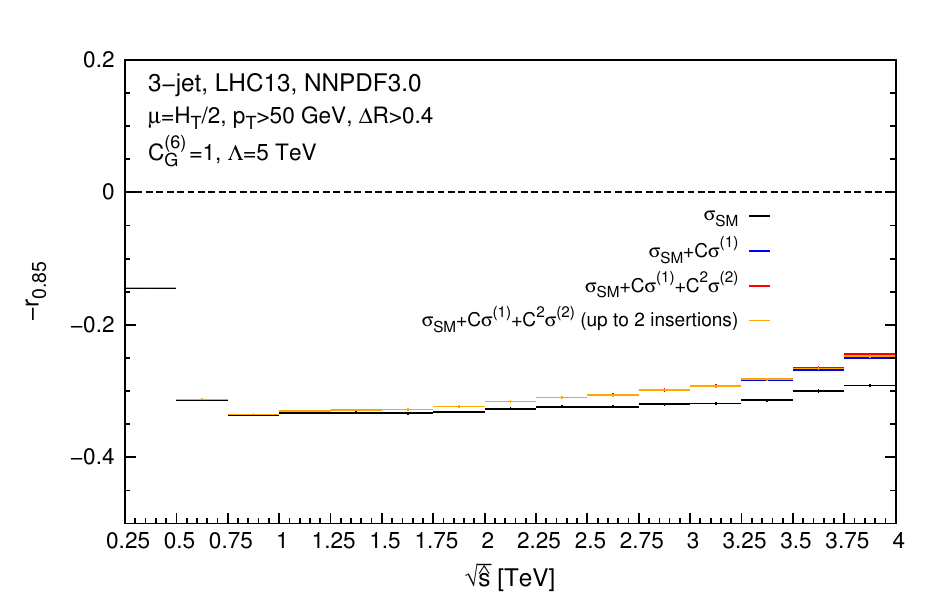}
\includegraphics[width=0.6\textwidth,trim=0cm 0cm 0cm 0cm]{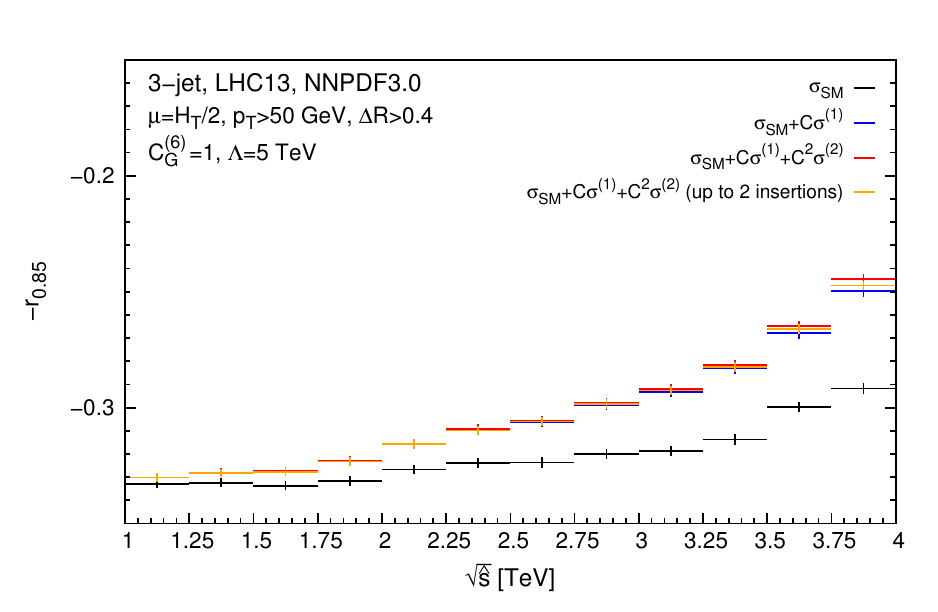}
\caption{\label{figure5Dixonzoom} Distribution of $r_{0.85}$ as defined in Eq.~\ref{rdefDixon} as a function of the c.o.m energy in three-jet production at LHC13 with the cuts defined Eq.~\ref{setCcuts}. The lower plot zooms in the high-energy region of the upper one.}
\end{center}
\end{figure}

It is well understood that the quantities entering in this observable are not invariant under boost along the beam axis and, as such, they are very challenging to reconstruct in hadronic collisions. We leave the determination of experimentally-viable proxies to these quantities to future work, and focus here only on investigating the potential of the observable $r_{0.85}$ at LHC13, in an ideal scenario where it can be perfectly measured. We compute it at the LHC with the additional \emph{modern} cuts of Eq.~\ref{setCcutsadd}, hence insuring that the phase-space region considered is at least experimentally accessible. In this case, we extend the range to 4 TeV and consider the case of $C_G=1$ and $\Lambda=5$ TeV so as to match the limit currently placed from the multi-jet analysis. We present our results in Figure~\ref{figure5Dixonzoom}, for the contributions of order $\mathcal{O}(1/\Lambda^2)$ and $\mathcal{O}(1/\Lambda^4)$, with up to either one or two insertions of the $O_G$ operator in the contributing amplitudes. 

We find that the contribution linear in $\mathcal{O}(1/\Lambda^2)$ dominates and yields a significant deviation from the SM. This is a rather unique example of an observable for which the $O_G$ signal is dominated by the linear contribution at a high energy and also of significant size compared to the SM prediction (up to $\sim$15\% in the tail). Furthermore, $r_{0.85}$ being a ratio observable, it is prone to cancellation of the correlated theoretical uncertainties and therefore offers good prospects for constraining the $O_G$ operator. We verify this assumption by computing the LO renormalisation scale uncertainties which are indeed found to be very moderate for this observable\footnote{Our study is based on LO parton-level simulations and therefore our conclusions come with the caveat that the transition towards the collinear region could be significantly affected by the resummation/parton shower corrections that are ignored here.}.
The viability of this observable therefore crucially depends on the ability to reconstruct the peculiar quantities entering the definition of $r_{0.85}$ and on the experimental accuracy that can be achieved in the contrived region of phase-space considered here.

\begin{figure}[t]
\begin{center}
\includegraphics[width=0.6\textwidth,trim=0cm 0cm 0cm 0cm]{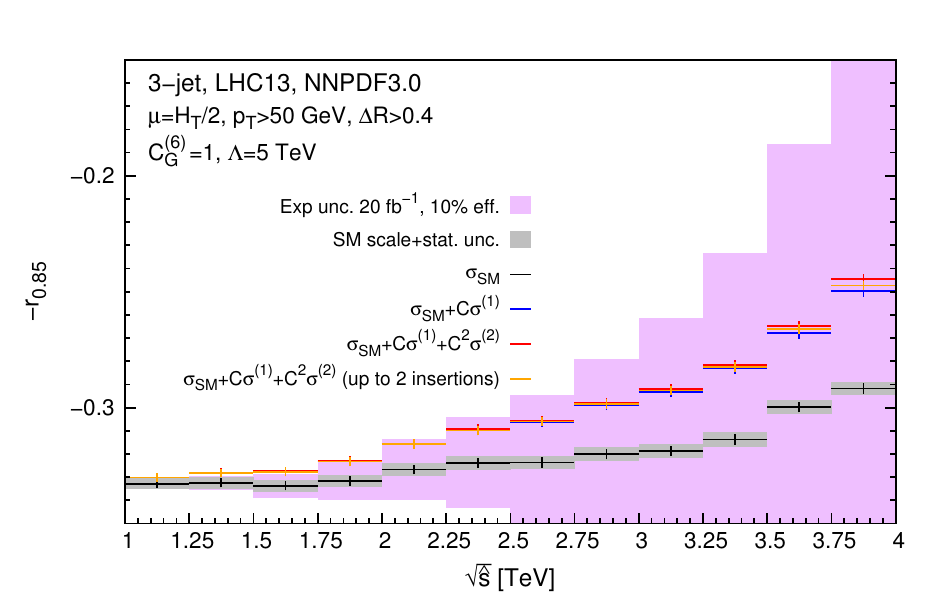}
\includegraphics[width=0.6\textwidth,trim=0cm 0cm 0cm 0cm]{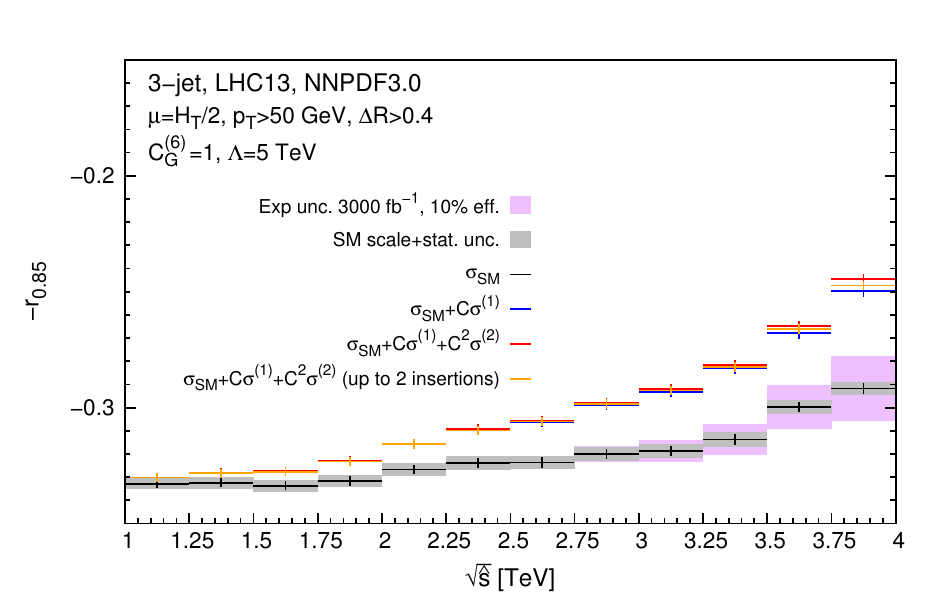}
\caption{\label{figure5DixonzoomLumin} Distribution of $r_{0.85}$ as defined in Equation as a function of the centre-of-mass energy in three-jet production with the cuts defined in Eqs.~\ref{setCcuts} and~\ref{setCcutsadd}, along with the expected experimental uncertainties for different integrated luminosities at the LHC. }
\end{center}
\end{figure}

We elaborate on the latter of these requirements, by assuming that the experimental uncertainty is entirely of statistical nature and compute it from the expected number of events in each bin of the $x_3$ distribution entering in the definition of Eq.~\ref{rdefDixon}, multiplied by an acceptance efficiency of 10\%, which we believe to be a realistic estimate of the efficiency, given the complexity of this observable. The experimental errors hence obtained for each of the four terms of Eq.~\ref{rdefDixon} are then propagated to yield the one of the ratio-observable $r_{0.85}$, as shown in Figure~\ref{figure5DixonzoomLumin}. We considered an integrated luminosity of both 20 fb$^{-1}$, relevant for current measurements, and 3000 fb$^{-1}$, as in the High Luminosity LHC (HL-LHC) scenario. We find that even with the current integrated luminosity, the first few  bins in the range between 1 and 2 TeV lie close to the edge of the experimental uncertainty band, therefore indicating that this observable could in principle already be used to place competitive constraints on the $O_G$ operator. For the HL-LHC scenario, the situation is even more promising as the statistics of the high-energy bins featuring the larger deviation w.r.t the SM is large enough so as to potentially constrain the $O_G$ operator beyond the current limits.

We stress again that our study is a simplified one of an idealised observable, with various aspects requiring a more realistic treatment. In particular, the quantities entering in the definition of $r_{0.85}$ are expected to be especially difficult to reconstruct in hadronic collisions. The ordering of the jets and the cuts applied in the c.o.m. frame should be traded for an observable more easily reconstructed than $x_3$ but capturing the same physics. Another shortcoming of our analysis is that we did not consider matching to parton showers, which could potentially modify the outcomes. Finally our estimate of the experimental and theoretical uncertainties is very simplified, and calls for a more careful investigation in order to formulate a definite statement. We therefore restrict our conclusion here to stating that the idealised observable $r_{0.85}$ of Eq.~\ref{rdefDixon}, originally proposed in~\cite{Dixon:1993xd}, retains its discriminatory power at LHC13. This opens the possibility of deriving a more realistic analysis that could place a strong limit on the $O_G$ operator, solely from the interference terms of order $\mathcal{O}(1/\Lambda^2)$, thereby rendering it more robust from the EFT-validity standpoint.

\section{Conclusions and Outlook}
We have examined the impact of the effective triple gluon operator $O_G$ on a series of observables at the LHC and in the light of the limit of $\Lambda=5$ TeV recently placed on its coefficient using high-energy multi-jet events. Focussing on the same observable, i.e. the transverse energy of multi-jet events, we have investigated the dependence of the reach of the analysis on the jet multiplicity, concluding that the opening of new partonic channels featuring enhanced $O_G$ contributions w.r.t the SM is ultimately responsible for the increased sensitivity. We also confirmed that the limit hence obtained is valid within the EFT expansion by computing the impact of dimension-8 operators on the same observable. 


We then investigated the impact of $O_G$ in various other jet observables, in particular in three-jet and four-jet events with well separated and hard jets. We found that the signal is most often dominated by terms of order $\mathcal{O}(1/\Lambda^4)$ while the interference contribution of order $\mathcal{O}(1/\Lambda^2)$ is only relevant in cases where the overall signal strength is below the percent level. Similarly, we showed that heavy-quark pair production processes lead to very small deviations w.r.t to the SM and are unlikely to improve the limits on this operator. 

Given the stringent constraints that can be set on the triple gluon operator using high-jet multiplicity  measurements one might also wonder whether the CP-odd operator $O_{\tilde G}$ can also be constrained from the same measurements.  On the one hand,  it is well known that $O_{\tilde G}$ is strongly bound  by the neutron EDM, see for example \cite{Dekens:2013zca}, giving a limit on $\Lambda$ for $C_{\tilde{G}}(M_W)=1$ of about 30 TeV.  On the other hand,  these limits suffer from model assumptions as they are indirect. In this respect, even though weaker, collider constraints could provide complementary information. It is straightforward to evaluate the impact of $O_{\tilde G}$ on the variable $S_T$, which is found to be the same as that of $O_{G}$ in the high $S_T$ region which is the one setting the limits, therefore yielding a bound of $\Lambda> 5$ TeV. Whilst this is a naive estimate of a potential limit, it shows that collider results can potentially compete with indirect bounds as higher energies get probed and more studies on $O_{\tilde G}$ could be welcome.

We have also computed for the first time here the contribution of $O_G$ to di-jet at next-to-leading order in QCD. Di-jet production receives $O_G$ interference contributions of order $\mathcal{O}(1/\Lambda^2)$ when considering one-loop amplitudes featuring up to one insertion of the $O_G$ operator. We established that they have no phenomenological relevance as they are either too small compared the SM yield or trumped by the contributions of order $\mathcal{O}(1/\Lambda^4)$, depending on the jet $p_T$ range considered.

Finally, we have revisited in the context of the LHC a particular three-jet angular observable first suggested in~\cite{Dixon:1993xd} for constraining the triple gluon operator at the Tevatron. This ratio-observable, which is rather complicated,  capitalises on the peculiar behaviour of $O_G$ operator in the collinear limit and we show that it retains its ability to discriminate against both QCD and $\mathcal{O}(1/\Lambda^4)$ contributions at the LHC, even with the stringent multi-jet limits. Moreover, our very preliminary analysis of theoretical and statistical uncertainty indicates that LHC jet data could potentially further constrain this operator, solely from its linear contributions which are more robust from an EFT expansion standpoint. While our analysis requires further investigation, we find promising results that will hopefully prompt additional studies on the topic.


\section{Acknowledgements}

VH wishes to thank C\'eline Degrande for her help with the use of {\sc NLOCT} and Lance Dixon for his comments on the draft of this publication as well as his guidance on the analytical understanding of the structure of the one-loop $O_G$ amplitudes discussed in appendix~\ref{analyticGGG}. VH also acknowledges SLAC computing resources. The work of VH is supported by the ERC grant 694712 ``pertQCD''. FM is partially supported by the  F.R.S.-FNRS under the `Excellence of Science' EOS be.h project n. 30820817. The work of IT is supported by the F.R.S.-FNRS `Fonds de la Recherche Scientifique' (Belgium). IT also thanks the department of physics at TUM (Munich) for the hospitality during the completion of this project. EV is supported by a Marie Sk\l{}odowska-Curie Individual Fellowship of the European Commission's Horizon 2020 Programme under contract number 704187.

\appendix

\section{Analytic one-loop matrix elements for $g g \rightarrow g g$ at order $\mathcal{O}(1/\Lambda^2)$}
\label{analyticGGG}

The one-loop computations necessary for this work were performed using the automated and numerical toolchain {\sc FeynRules}~\cite{Alloul:2013bka}/{\sc NLOCT}~\cite{Degrande:2014vpa} + {\sc MadLoop}~\cite{Hirschi:2011pa}/{\sc MG5aMC}~\cite{Alwall:2014hca,Hirschi:2015iia}. This head-on approach is typically less error-prone and allows to easily include all partonic channels as well as marginal effects such as including the top-quark mass dependence in the loop. It also provides a straight-forward access to the flexibility offered by event generators.
 
However, given the simplicity of the $2 \rightarrow 2$ kinematics of the dijet amplitudes relevant here, it is both possible and interesting to peek at the analytical structure of the loop amplitudes computed in this work for the first time. In this appendix, we therefore explore in more analytical depth the various contributions of the operator $O_G$ of Eq.~\ref{eq:3G} to the four-gluon amplitude. In the following we use the well-known spinor helicity notation, see Refs.~\cite{Mangano:1990by,Dixon:2013uaa}. We introduce first  the tree-level amplitudes of the process $g g \rightarrow g g$ and then investigate the contribution driven by the known one-loop QCD amplitude for that process. Finally, we share our original analytic result for the one-loop four-gluon amplitude featuring one insertion of the $O_G$ operator and obtained via a numerical fit of \MadLoop's result.

Multi-gluon amplitudes at the tree level can be written in terms of partial amplitudes $\mathcal{A}$ as
\begin{eqnarray}
\label{dualdecomposition}
\mathcal{M}_{\tree}(\{p_i\}_1^n, \{\epsilon_i\}_1^n, \{a_i\}_1^n)=\sum_{\sigma_I\in P(\{2,\cdots,n)\}} \big{[}\; \Tr(t^{a_1}t^{a_{\sigma_I(2)}}\cdots t^{a_{\sigma_I(n)}}) \nonumber\\
\mathcal{A}_n(p_1, \epsilon_1; p_{\sigma_I(2)},\epsilon_{\sigma_I(2)}; \cdots ; p_{\sigma_I(n)},\epsilon_{\sigma_I(n)}) \;\big{]},
\end{eqnarray}
where $\{p_i\}_1^n$, $\{\epsilon_i\}_1^n$ and $\{a_i\}_1^n$ denote the set of $n$ momenta, polarisation vectors and adjoint colour indices of the $n$ gluons considered. 

The four gluon partial amplitudes $\mathcal{A}^{(\QCD)}_{4,\tree}$, read~\cite{Parke:1986gb}:
\begin{eqnarray}
\label{ggggDualAmp}
&&\mathcal{A}^{(\QCD)}_{4,\tree}(i^+,j^+,k^+,l^+)=0\\
&&\mathcal{A}^{(\QCD)}_{4,\tree}(i^-,j^+,k^+,l^+)=0\\
&&\mathcal{A}^{(\QCD)}_{4,\tree}(i^-,j^-,k^+,l^+)=\mi g_s^2 \frac{\bk{ij}^4}{\bk{ij}\bk{jk}\bk{kl}\bk{li}}\\
&&\mathcal{A}^{(\QCD)}_{4,\tree}(i^-,j^+,k^-,l^+)=\mi g_s^2 \frac{\bk{ik}^4}{\bk{ij}\bk{jk}\bk{kl}\bk{li}}\,,
\end{eqnarray}
where $g_s$ denotes the strong coupling constant. The tree-level four-gluon partial amplitudes with one insertion of the $O_G$ operator have been computed in ref.~\cite{Dixon:1993xd}, which we denote $\mathcal{A}^{(\Lambda)}_{4,\tree}$ and read:
\begin{eqnarray}
\label{GGGtree}
&&\mathcal{A}^{(\Lambda)}_{4,\tree}(1^+,2^+,3^+,4^+)=
g_s^2 \frac{3 \mi}{\Lambda^2}\frac{2 s t u}{\bk{12}\bk{23}\bk{34}\bk{41}}\nonumber\\
&&\mathcal{A}^{(\Lambda)}_{4,\tree}(1^-,2^+,3^+,4^+)=
-g_s^2\frac{3 \mi}{\Lambda^2}\frac{\kk{23}^2\kk{34}^2\kk{42}^2}{\kk{12}\kk{23}\kk{34}\kk{41}}
\nonumber\\
&&\mathcal{A}^{(\Lambda)}_{4,\tree}(1^-,2^-,3^+,4^+)=0\nonumber\\
&&\mathcal{A}^{(\Lambda)}_{4,\tree}(1^-,2^+,3^-,4^+)=0,
\end{eqnarray}
where we used consecutive integer numbers in the argument of the partial amplitudes to label external particles. We stress that the results presented here are however valid for any assignment of external particle labels, although the reader must keep in mind that the usual Mandelstam variables $s$, $t$ and $u$ (with $s+t+u=0$)  are bound to a particular assignment. The results above show why the interference contribution $M^{(\Lambda)\times \QCD}_{\tree \times \tree}$ is exactly zero at the tree level: the helicity configuration supports of the $\QCD$ and $\EFT$ partial amplitudes are orthogonal to each other (see Eqs.~\ref{ggggDualAmp} and~\ref{GGGtree}).
At \emph{tree-level}, the $O_G$ operator therefore only contributes to the four-gluon matrix element via $M^{(\Lambda)\times (\Lambda)}_{\tree \times \tree}$, which is of order $\mathcal{O}(1/\Lambda^4)$. In order to unlock the linear $O_G$ contribution of order $\mathcal{O}(1/\Lambda^2)$, one must open up new helicity configurations, for example by considering one-loop contributions to the four-gluon amplitudes. The $O_G$ operator can then be inserted either in the tree-amplitude multiplying the one-loop QCD one, yielding the contribution denoted with $M^{(\QCD)\times (\Lambda)}_{\looptxt \times \tree}$ or directly in the loop, yielding $M^{(\Lambda)\times (\QCD)}_{\looptxt \times \tree}$. In all the results presented here, the coupling $C_G$ is set to one.

We now present the QCD one-loop partial amplitudes $\mathcal{A}^{\QCD}_{4,\looptxt}$, as given in ref.~\cite{Bern:1991aq}. The colour basis introduced in Eq.~\ref{dualdecomposition} must be expanded in order to be able to project one-loop amplitudes, and the resulting decomposition can be generalised as follows:
\begin{eqnarray}
\label{loopdualdecomposition}
\mathcal{M}_{\looptxt}(\{p_i\}_1^n, \{\epsilon_i\}_1^n, \{a_i\}_1^n)=
\sum_{j=1}^{[n/2]+1}\sum_{\sigma_I\in S_n / S_{n;j}} \big{[}\; \Gr_{n;j} (\sigma_I(1),...,\sigma_I(n)) \nonumber\\
\mathcal{A}_{n;j}(\sigma_I(1),...,\sigma_I(n)) \;\big{]}
\end{eqnarray}
with:
\begin{eqnarray}
&&\Gr_{n;1}(1,...,n)= \Tr(\identity)\Tr(t^{a_1},...,t^{a_n}) = N_c\Tr(t^{a_1},...,t^{a_n}) \nonumber\\
&&\Gr_{n;j}(1,...,n)=\Tr(t^{a_1},...,t^{a_{j-1}})\Tr(t^{a_j},...,t^{a_n}).
\end{eqnarray}
In Eq.~\ref{loopdualdecomposition}, the cyclic symmetry of the color traces is no longer lifted by setting their first indices and it is therefore modded out from the permutation group summed over. Indeed, we define permutations as part of the group $S_n$ of all permutations of $n$ indices, removed of all permutations in $S_{n;j}$ which, by definition, leave the double color trace $\Gr_{n;j}$ invariant up to their respective cyclic symmetry. A  more compact notation based on the colour adjoint representation \cite{DelDuca:1999rs} is sufficient to show that in fact only the leading colour amplitudes $\mathcal{A}_{n;1}(\sigma_I(1),...,\sigma_I(n))$ are truly independent. 
One can then compactly write the colour-summed loop matrix element (consisting of loop and tree amplitudes interferences) for $gg\to gg$ in terms of $\mathcal{A}_{4;1}$ only, i.e.
\begin{equation}
\label{NLOME}
M_{\looptxt \times \tree}^{(4g)} (\{i^\pm\}_1^n) = N_c^3 (N_c^2-1) \sum_{\sigma \in S_n/S_{n;1}} 2 \Re\left( \mathcal{A}_{4;1}^{\looptxt}(\sigma(\{i^\pm\}_1^n)) \mathcal{A}_4^{\tree}(\sigma(\{i^\pm\}_1^n))^{\star}\right).
\end{equation}

The explicit analytical expressions of the QCD 1-loop partial amplitudes $\mathcal{A}_{4;1}^{\QCD}$ then read (see ref.~\cite{Bern:1990ux}):
\begin{eqnarray}
\label{QCDloop}
&& \mathcal{A}_{4;1}^{\QCD} (1^+, 2^+, 3^+, 4^+) = \frac{\mi g_s^4}{48\pi^2} \frac{s t}{\bk{12}\bk{23}\bk{34}\bk{41}} \nonumber\\
&& \mathcal{A}_{4;1}^{\QCD} (1^-, 2^+, 3^+, 4^+) = \frac{\mi g_s^4}{48\pi^2} \frac{\kk{24}^2(t+s)}{[12]\bk{23}\bk{34}[41]}.
\end{eqnarray}
Substituting Eqs.~\ref{QCDloop} and~\ref{GGGtree} in Eq.~\ref{NLOME} then yields the exact expression for the pure gluon contribution in $M_{\looptxt \times \tree}^{\QCD \times \EFT}$. The all-minus and all-plus helicity contributions vanish, being proportional to $s+t+u$, as one would expect from their symmetries. 
For the helicity configuration $-+++$, we find:
\begin{equation}
M(1^-,2^+,3^+,4^+) =2 N_c^3(N_c^2-1) \frac{g_s^6}{8\pi^2}  \Lambda^{-2} \left(-1\right)\frac{s^4+t^4+u^4}{s t u}.
\end{equation}
One then arrives at the final result by summing over the eight contributions from all cyclic and CP permutations of the result above:
\begin{equation}
\label{QCDloopXGGGtreeResult}
M_{\looptxt \times \tree}^{\QCD \times \EFT} (g g \rightarrow g g) |_{n_f=0}=\frac{1}{2}\frac{1}{2\cdot2\cdot8\cdot8}  N_c^3 (N_c^2-1) \Lambda^{-2}\frac{g_s^6}{8 \pi^2}(16)\frac{s^4+t^4+u^4}{stu}.
\end{equation}
We note that we checked numerically that including $n_f$ massless fermion flavours affects the result above by an overall factor $(1-\frac{n_f}{N_c})$, as it could be anticipated from the supersymmetric case where the physical contribution vanishes when considering an equal number of colour-charged bosonic and fermionic degrees of freedom (of identical mass).

We can now consider the loop amplitudes in the EFT theory. As previously stated, the $\mathcal{A}_{\looptxt}^{\EFT}$ amplitudes have been computed for the first time and numerically in our work. Taking advantage of \MadLoop's capability of providing numerical evaluations for specific helicity and colour-flow configurations, we could derive the analytical expression of the amplitudes $\mathcal{A}_{4;j}^{\EFT}$ using a numerical fitting procedure on an ansatz factorising the QCD partial amplitudes $\mathcal{A}^{(\QCD)}_{4,\tree}$ in order to remove the dependency on the complex phase introduced by the specific choice of polarisation vectors used by \MadLoop. We obtain:
\begin{eqnarray}
\label{GGGloop}
&& \mathcal{A}_{4;1}^{\EFT} (1^-, 2^-, 3^+, 4^+) = \left(-(3+\frac{n_f}{2})t+(3-n_f)\frac{u t}{s}\right)\Lambda^{-2}\frac{g_s^2}{8\pi^2}\mathcal{A}^{(\QCD)}_{4,\tree}(1^-,2^-,3^+,4^+) \nonumber\\
&& \mathcal{A}_{4;1}^{\EFT} (1^-, 2^+, 3^-, 4^+) = 0,
\end{eqnarray}
where $N_c$ has now been explicitly set to 3.
We have verified the cyclic symmetry of the partial amplitudes obtained from our numerical fitting procedure. Although not necessary for the computation of the physical contribution $M_{\looptxt \times \tree}^{\EFT \times \QCD}$, we also provide here the analytical expression for the partial amplitudes $\mathcal{A}_{4;3}^{\EFT}$, obtained using a numerical fit as well:
\begin{eqnarray}
\label{GGGloop43}
&& \mathcal{A}_{4;3}^{\EFT} (1^-, 2^-, 3^+, 4^+) = \left(-6 t \right)\Lambda^{-2}\frac{g_s^2}{8\pi^2}\mathcal{A}^{(\QCD)}_{4,\tree}(1^-,2^-,3^+,4^+) \nonumber\\
&& \mathcal{A}_{4;3}^{\EFT} (1^-, 2^+, 3^-, 4^+) = \left(-6 \frac{s t}{u} \right)\Lambda^{-2}\frac{g_s^2}{8\pi^2}\mathcal{A}^{(\QCD)}_{4,\tree}(1^-,2^+,3^-,4^+).
\end{eqnarray}
Summing over all non-cyclic permutations of the partial amplitudes $\mathcal{A}_{4;1}$ for the helicity configuration $--++$ and $-+-+$, we find their contributions to be:
\begin{eqnarray}
M(1^-,2^-,3^+,4^+) = 2\cdot 3^3(3^2-1) \frac{g_s^6}{8\pi^2} \Lambda^{-2}\left( \frac{t+u}{u t}((n_f-12)(t^2+u^2) - 2(n_f+6) t u) \right) \nonumber\\
M(1^-,2^+,3^-,4^+) = 2\cdot 3^3(3^2-1) \frac{g_s^6}{8\pi^2} \Lambda^{-2}\left( \frac{t+s}{s t}((n_f-12)(t^2+s^2) - 2(n_f+6) t s) \right). \nonumber\\
\end{eqnarray}
Finally summing over all helicity contributions, we arrive at the final expression for $M_{\looptxt \times \tree}^{\EFT \times \QCD}$:
\begin{equation}
M_{\looptxt \times \tree}^{\EFT \times \QCD} (g g \rightarrow g g)=\frac{1}{2}\frac{1}{2\cdot2\cdot8\cdot8}  3^3(3^2-1) \Lambda^{-2}\frac{g_s^6}{8 \pi^2}\left(16\left(1-\frac{n_f}{12}\right)\right)\frac{s^4+t^4+u^4}{stu}.
\end{equation}
It is interesting to note that for the purely gluonic contribution ($n_f=0$), we have:
\begin{equation}
M_{\looptxt \times \tree}^{\EFT \times \QCD} (g g \rightarrow g g) |_{n_f=0} = M_{\looptxt \times \tree}^{\QCD \times \EFT} (g g \rightarrow g g) |_{n_f=0}\,,
\end{equation}
while for six massless fermions ($n_f=6$):
\begin{equation}
M_{\looptxt \times \tree}^{\EFT \times \QCD} (g g \rightarrow g g) |_{n_f=6} = \left( -\frac{1}{2} \right) M_{\looptxt \times \tree}^{\QCD \times \EFT} (g g \rightarrow g g) |_{n_f=6}.
\end{equation}
The rational one-loop finite four-gluon matrix element summed over all colours and helicity configurations must be completely symmetric upon any relabelling of the external gluons. This greatly restricts the functional space in which it can be expressed and perhaps renders the above association less of a surprise. Indeed, when considering channels with external quark lines or introduce new scales, in the form of the top-quark mass in the loop for example, then the analytical treatment presented in this appendix quickly becomes more cumbersome and the two contributions $M_{\looptxt \times \tree}^{\EFT \times \QCD}$ and $M_{\looptxt \times \tree}^{\QCD \times \EFT}$ exhibit different functional dependence. In such cases, we resort only to the numerical computation performed by \MadLoop.

\section{Partonic channel decomposition in multi-jet cross section}
\label{app:tables}
In this appendix we show  in Tables \ref{table:2-jet}-\ref{table:4-jet} the contributions of each partonic channel to the dijet, three-jet and four-jet final states as a function of the cut on $S_T$. The relative contribution to the total cross section and the impact of $O_G$ compared to the SM prediction is also shown for all partonic channels.
\begin{table}[H]
\renewcommand{\arraystretch}{1.25}
\scriptsize
\begin{center}
\begin{tabular}{c | c c || c | c c || c}
\hline
\hline
\multicolumn{7}{c}{$p p \rightarrow j j $} \\
\hline
\multicolumn{3}{c ||}{SM} & \multicolumn{3}{c ||}{SM + $O_G [\mathcal{O}(1/\Lambda^2,1/\Lambda^4)]$} & \multirow{2}{*}{(SM + $O_G$)/SM}  \\
\cline{1-6}
Subprocess	&	$\sigma$ [pb]	&	fraction [\%]	&	Subprocess	&	$\sigma$ [pb]	&	fraction [\%]	&	\\
\hline													
$gg \rightarrow gg$	&	1.11$\cdot 10^{+7}$	&	54.97	&	$gg \rightarrow gg$	&	1.11$\cdot 10^{+7}$	&	54.97	&	1.00	\\
$gg \rightarrow qq$	&	3.93$\cdot 10^{+5}$	&	1.94	&	$gg \rightarrow qq$	&	3.93$\cdot 10^{+5}$	&	1.94	&	1.00	\\
$qg \rightarrow qg$	&	7.72$\cdot 10^{+6}$	&	38.16	&	$qg \rightarrow qg$	&	7.72$\cdot 10^{+6}$	&	38.16	&	1.00	\\
$qq \rightarrow gg$	&	1.35$\cdot 10^{+4}$	&	0.07	&	$qq \rightarrow gg$	&	1.35$\cdot 10^{+4}$	&	0.07	&	1.00	\\
$qq \rightarrow qq$	&	9.83$\cdot 10^{+5}$	&	4.86	&	$qq \rightarrow qq$	&	9.83$\cdot 10^{+5}$	&	4.86	&	1.00	\\
total	&	2.02$\cdot 10^{+7}$	&	100.00	&	total	&	2.02$\cdot 10^{+7}$	&	100.00	&	1.00	\\
\hline
\hline
\multicolumn{7}{c}{$S_T > 2$ TeV} \\
\hline
\multicolumn{3}{c ||}{SM} & \multicolumn{3}{c ||}{SM + $O_G [\mathcal{O}(1/\Lambda^2,1/\Lambda^4)]$} & \multirow{2}{*}{(SM + $O_G$)/SM}  \\
\cline{1-6}
Subprocess	&	$\sigma$ [pb]	&	fraction [\%]	&	Subprocess	&	$\sigma$ [pb]	&	fraction [\%]	&		\\
\hline													
$gg \rightarrow gg$	&	8.56$\cdot 10^{-1}$	&	9.91	&	$gg \rightarrow gg$	&	9.99$\cdot 10^{-1}$	&	10.98	&	1.17	\\
$gg \rightarrow qq$	&	3.83$\cdot 10^{-2}$	&	0.44	&	$gg \rightarrow qq$	&	5.47$\cdot 10^{-2}$	&	0.60	&	1.43	\\
$qg \rightarrow qg$	&	4.44$\cdot 10^{0}$	&	51.37	&	$qg \rightarrow qg$	&	4.72$\cdot 10^{0}$	&	51.83	&	1.06	\\
$qq \rightarrow gg$	&	4.35$\cdot 10^{-2}$	&	0.50	&	$qq \rightarrow gg$	&	6.49$\cdot 10^{-2}$	&	0.71	&	1.49	\\
$qq \rightarrow qq$	&	3.27$\cdot 10^{0}$	&	37.78	&	$qq \rightarrow qq$	&	3.27$\cdot 10^{0}$	&	35.88	&	1.00	\\
total	&	8.64$\cdot 10^{0}$	&	100.00	&	total	&	9.10$\cdot 10^{0}$	&	100.00	&	1.05	\\
\hline
\hline
\multicolumn{7}{c}{$S_T > 3$ TeV} \\
\hline
\multicolumn{3}{c ||}{SM} & \multicolumn{3}{c ||}{SM + $O_G [\mathcal{O}(1/\Lambda^2,1/\Lambda^4)]$} & \multirow{2}{*}{(SM + $O_G$)/SM}  \\
\cline{1-6}
Subprocess	&	$\sigma$ [pb]	&	fraction [\%]	&	Subprocess	&	$\sigma$ [pb]	&	fraction [\%]	&		\\
\hline													
$gg \rightarrow gg$	&	2.04$\cdot 10^{-2}$	&	4.34	&	$gg \rightarrow gg$	&	3.55$\cdot 10^{-2}$	&	6.50	&	1.74	\\
$gg \rightarrow qq$	&	9.48$\cdot 10^{-4}$	&	0.20	&	$gg \rightarrow qq$	&	2.75$\cdot 10^{-3}$	&	0.50	&	2.90	\\
$qg \rightarrow qg$	&	1.96$\cdot 10^{-1}$	&	41.67	&	$qg \rightarrow qg$	&	2.50$\cdot 10^{-1}$	&	45.74	&	1.28	\\
$qq \rightarrow gg$	&	2.49$\cdot 10^{-3}$	&	0.53	&	$qq \rightarrow gg$	&	7.89$\cdot 10^{-3}$	&	1.44	&	3.16	\\
$qq \rightarrow qq$	&	2.51$\cdot 10^{-1}$	&	53.26	&	$qq \rightarrow qq$	&	2.51$\cdot 10^{-1}$	&	45.82	&	1.00	\\
total	&	4.70$\cdot 10^{-1}$	&	100.00	&	total	&	5.47$\cdot 10^{-1}$	&	100.00	&	1.16	\\
\hline
\hline
\multicolumn{7}{c}{$S_T > 4$ TeV} \\
\hline
\multicolumn{3}{c ||}{SM} & \multicolumn{3}{c ||}{SM + $O_G [\mathcal{O}(1/\Lambda^2,1/\Lambda^4)]$} & \multirow{2}{*}{(SM + $O_G$)/SM}  \\
\cline{1-6}
Subprocess	&	$\sigma$ [pb]	&	fraction [\%]	&	Subprocess	&	$\sigma$ [pb]	&	fraction [\%]	&		\\
\hline												
$gg \rightarrow gg$	&	7.01$\cdot 10^{-4}$	&	1.73	&	$gg \rightarrow gg$	&	2.20$\cdot 10^{-3}$	&	4.12	&	3.14	\\
$gg \rightarrow qq$	&	3.37$\cdot 10^{-5}$	&	0.08	&	$gg \rightarrow qq$	&	2.21$\cdot 10^{-4}$	&	0.41	&	6.56	\\
$qg \rightarrow qg$	&	1.25$\cdot 10^{-2}$	&	31.02	&	$qg \rightarrow qg$	&	2.25$\cdot 10^{-2}$	&	42.13	&	1.80	\\
$qq \rightarrow gg$	&	2.16$\cdot 10^{-4}$	&	0.53	&	$qq \rightarrow gg$	&	1.58$\cdot 10^{-3}$	&	2.96	&	7.35	\\
$qq \rightarrow qq$	&	2.69$\cdot 10^{-2}$	&	66.63	&	$qq \rightarrow qq$	&	2.69$\cdot 10^{-2}$	&	50.38	&	1.00	\\
total	&	4.04$\cdot 10^{-2}$	&	100.00	&	total	&	5.35$\cdot 10^{-2}$	&	100.00	&	1.32	\\
\hline
\end{tabular}
\end{center}
 \caption{\label{table:2-jet} Contributions from the different partonic subprocesses contributing to di-jet production at LHC13 in the SM ($\mathcal{O}(1/\Lambda^0)$) and when allowing at most one $O_G$ insertion in the amplitudes ($\mathcal{O}(1/\Lambda^2)$ and $\mathcal{O}(1/\Lambda^4)$). }  
\end{table}

\begin{table}[t]
\renewcommand{\arraystretch}{1.25}
\scriptsize
\begin{center}
\begin{tabular}{c | c c || c | c c || c}
\hline
\hline
\multicolumn{7}{c}{$p p \rightarrow j j j$} \\
\hline
\multicolumn{3}{c ||}{SM} & \multicolumn{3}{c ||}{SM + $O_G [\mathcal{O}(1/\Lambda^2,1/\Lambda^4)]$} & \multirow{2}{*}{(SM + $O_G$)/SM}  \\
\cline{1-6}
Subprocess	&	$\sigma$ [pb]	&	fraction [\%]	&	Subprocess	&	$\sigma$ [pb]	&	fraction [\%]	&	\\
\hline													
$gg \rightarrow ggg$	&	3.79$\cdot 10^{+5}$	&	42.88	&	$gg \rightarrow ggg$	&	3.79$\cdot 10^{+5}$	&	42.88	&	1.00	\\
$gg \rightarrow gqq$	&	4.46$\cdot 10^{+4}$	&	5.04	&	$gg \rightarrow gqq$	&	4.46$\cdot 10^{+4}$	&	5.04	&	1.00	\\
$qg \rightarrow ggq$	&	3.70$\cdot 10^{+5}$	&	41.85	&	$qg \rightarrow ggq$	&	3.70$\cdot 10^{+5}$	&	41.85	&	1.00	\\
$qg \rightarrow qqq$	&	2.07$\cdot 10^{+4}$	&	2.34	&	$qg \rightarrow qqq$	&	2.07$\cdot 10^{+4}$	&	2.34	&	1.00	\\
$qq \rightarrow ggg$	&	3.74$\cdot 10^{+2}$	&	0.04	&	$qq \rightarrow ggg$	&	3.74$\cdot 10^{+2}$	&	0.04	&	1.00	\\
$qq \rightarrow gqq$	&	6.95$\cdot 10^{+4}$	&	7.85	&	$qq \rightarrow gqq$	&	6.95$\cdot 10^{+4}$	&	7.85	&	1.00	\\
total	&	8.85$\cdot 10^{+5}$	&	100.00	&	total	&	8.85$\cdot 10^{+5}$	&	100.00	&	1.00	\\
\hline
\hline
\multicolumn{7}{c}{$S_T > 2$ TeV} \\
\hline
\multicolumn{3}{c ||}{SM} & \multicolumn{3}{c ||}{SM + $O_G [\mathcal{O}(1/\Lambda^2,1/\Lambda^4)]$} & \multirow{2}{*}{(SM + $O_G$)/SM}  \\
\cline{1-6}
Subprocess	&	$\sigma$ [pb]	&	fraction [\%]	&	Subprocess	&	$\sigma$ [pb]	&	fraction [\%]	&		\\
\hline													
$gg \rightarrow ggg$	&	1.84$\cdot 10^{0}$	&	11.00	&	$gg \rightarrow ggg$	&	2.08$\cdot 10^{0}$	&	11.86	&	1.13	\\
$gg \rightarrow gqq$	&	2.02$\cdot 10^{-1}$	&	1.21	&	$gg \rightarrow gqq$	&	2.48$\cdot 10^{-1}$	&	1.41	&	1.23	\\
$qg \rightarrow ggq$	&	8.28$\cdot 10^{0}$	&	49.59	&	$qg \rightarrow ggq$	&	8.72$\cdot 10^{0}$	&	49.69	&	1.05	\\
$qg \rightarrow qqq$	&	4.41$\cdot 10^{-1}$	&	2.64	&	$qg \rightarrow qqq$	&	4.83$\cdot 10^{-1}$	&	2.75	&	1.10	\\
$qq \rightarrow ggg$	&	6.22$\cdot 10^{-2}$	&	0.37	&	$qq \rightarrow ggg$	&	9.24$\cdot 10^{-2}$	&	0.53	&	1.48	\\
$qq \rightarrow gqq$	&	5.87$\cdot 10^{0}$	&	35.19	&	$qq \rightarrow gqq$	&	5.92$\cdot 10^{0}$	&	33.75	&	1.01	\\
total	&	1.67$\cdot 10^{+1}$	&	100.00	&	total	&	1.75$\cdot 10^{+1}$	&	100.00	&	1.05	\\
\hline
\hline
\multicolumn{7}{c}{$S_T > 3$ TeV} \\
\hline
\multicolumn{3}{c ||}{SM} & \multicolumn{3}{c ||}{SM + $O_G [\mathcal{O}(1/\Lambda^2,1/\Lambda^4)]$} & \multirow{2}{*}{(SM + $O_G$)/SM}  \\
\cline{1-6}
Subprocess	&	$\sigma$ [pb]	&	fraction [\%]	&	Subprocess	&	$\sigma$ [pb]	&	fraction [\%]	&		\\
\hline													
$gg \rightarrow ggg$	&	4.52$\cdot 10^{-2}$	&	5.19	&	$gg \rightarrow ggg$	&	7.28$\cdot 10^{-2}$	&	7.06	&	1.61	\\
$gg \rightarrow gqq$	&	4.67$\cdot 10^{-3}$	&	0.54	&	$gg \rightarrow gqq$	&	9.60$\cdot 10^{-3}$	&	0.93	&	2.06	\\
$qg \rightarrow ggq$	&	3.70$\cdot 10^{-1}$	&	42.46	&	$qg \rightarrow ggq$	&	4.64$\cdot 10^{-1}$	&	45.04	&	1.25	\\
$qg \rightarrow qqq$	&	1.78$\cdot 10^{-2}$	&	2.04	&	$qg \rightarrow qqq$	&	2.44$\cdot 10^{-2}$	&	2.37	&	1.37	\\
$qq \rightarrow ggg$	&	4.02$\cdot 10^{-3}$	&	0.46	&	$qq \rightarrow ggg$	&	1.25$\cdot 10^{-2}$	&	1.21	&	3.11	\\
$qq \rightarrow gqq$	&	4.30$\cdot 10^{-1}$	&	49.32	&	$qq \rightarrow gqq$	&	4.47$\cdot 10^{-1}$	&	43.38	&	1.04	\\
total	&	8.71$\cdot 10^{-1}$	&	100.00	&	total	&	1.03$\cdot 10^{0}$	&	100.00	&	1.18	\\
\hline
\hline
\multicolumn{7}{c}{$S_T > 4$ TeV} \\
\hline
\multicolumn{3}{c ||}{SM} & \multicolumn{3}{c ||}{SM + $O_G [\mathcal{O}(1/\Lambda^2,1/\Lambda^4)]$} & \multirow{2}{*}{(SM + $O_G$)/SM}  \\
\cline{1-6}
Subprocess	&	$\sigma$ [pb]	&	fraction [\%]	&	Subprocess	&	$\sigma$ [pb]	&	fraction [\%]	&		\\
\hline												
$gg \rightarrow ggg$	&	1.52$\cdot 10^{-3}$	&	2.18	&	$gg \rightarrow ggg$	&	4.43$\cdot 10^{-3}$	&	4.46	&	2.91	\\
$gg \rightarrow gqq$	&	1.54$\cdot 10^{-4}$	&	0.22	&	$gg \rightarrow gqq$	&	6.36$\cdot 10^{-4}$	&	0.64	&	4.13	\\
$qg \rightarrow ggq$	&	2.33$\cdot 10^{-2}$	&	33.41	&	$qg \rightarrow ggq$	&	4.10$\cdot 10^{-2}$	&	41.35	&	1.76	\\
$qg \rightarrow qqq$	&	1.07$\cdot 10^{-3}$	&	1.54	&	$qg \rightarrow qqq$	&	2.20$\cdot 10^{-3}$	&	2.22	&	2.05	\\
$qq \rightarrow ggg$	&	3.56$\cdot 10^{-4}$	&	0.51	&	$qq \rightarrow ggg$	&	2.59$\cdot 10^{-3}$	&	2.61	&	7.28	\\
$qq \rightarrow gqq$	&	4.34$\cdot 10^{-2}$	&	62.15	&	$qq \rightarrow gqq$	&	4.84$\cdot 10^{-2}$	&	48.72	&	1.11	\\
total	&	6.99$\cdot 10^{-2}$	&	100.00	&	total	&	9.93$\cdot 10^{-2}$	&	100.00	&	1.42	\\
\hline
\end{tabular}
\end{center}
 \caption{\label{table:three-jet} 
 Contributions from the different partonic subprocesses contributing to the production of three partonic jets at LHC13 in the SM ($\mathcal{O}(1/\Lambda^0)$) and when allowing at most one $O_G$ insertion in the amplitudes ($\mathcal{O}(1/\Lambda^2)$ and $\mathcal{O}(1/\Lambda^4)$). }  
\end{table}

\begin{table}[!h]
\renewcommand{\arraystretch}{1.07}
\scriptsize
\begin{center}
\begin{tabular}{c | c c || c | c c || c}
\hline
\hline
\multicolumn{7}{c}{$p p \rightarrow j j j j$} \\
\hline
\multicolumn{3}{c ||}{SM} & \multicolumn{3}{c ||}{SM + $O_G [\mathcal{O}(1/\Lambda^2,1/\Lambda^4)]$} & \multirow{2}{*}{(SM + $O_G$)/SM}  \\
\cline{1-6}
Subprocess	&	$\sigma$ [pb]	&	fraction [\%]	&	Subprocess	&	$\sigma$ [pb]	&	fraction [\%]	&	\\
\hline													
$gg \rightarrow gggg$	&	4.27$\cdot 10^{+4}$	&	35.48	&	$gg \rightarrow gggg$	&	4.27$\cdot 10^{+4}$	&	35.47	&	1.00	\\
$gg \rightarrow ggqq$	&	9.05$\cdot 10^{+3}$	&	7.52	&	$gg \rightarrow ggqq$	&	9.05$\cdot 10^{+3}$	&	7.53	&	1.00	\\
$gg \rightarrow qqqq$	&	2.73$\cdot 10^{+2}$	&	0.23	&	$gg \rightarrow qqqq$	&	2.74$\cdot 10^{+2}$	&	0.23	&	1.00	\\
$qg \rightarrow gggq$	&	4.97$\cdot 10^{+4}$	&	41.35	&	$qg \rightarrow gggq$	&	4.97$\cdot 10^{+4}$	&	41.35	&	1.00	\\
$qg \rightarrow gqqq$	&	6.48$\cdot 10^{+3}$	&	5.38	&	$qg \rightarrow gqqq$	&	6.48$\cdot 10^{+3}$	&	5.39	&	1.00	\\
$qq \rightarrow gggg$	&	3.76$\cdot 10^{+1}$	&	0.03	&	$qq \rightarrow gggg$	&	3.76$\cdot 10^{+1}$	&	0.03	&	1.00	\\
$qq \rightarrow ggqq$	&	1.14$\cdot 10^{+4}$	&	9.50	&	$qq \rightarrow ggqq$	&	1.14$\cdot 10^{+4}$	&	9.50	&	1.00	\\
$qq \rightarrow qqqq$	&	6.05$\cdot 10^{+2}$	&	0.50	&	$qq \rightarrow qqqq$	&	6.05$\cdot 10^{+2}$	&	0.50	&	1.00	\\
total	&	1.20$\cdot 10^{+5}$	&	100.00	&	total	&	1.20$\cdot 10^{+5}$	&	100.00	&	1.00	\\
\hline
\hline
\multicolumn{7}{c}{$S_T > 2$ TeV} \\
\hline
\multicolumn{3}{c ||}{SM} & \multicolumn{3}{c ||}{SM + $O_G [\mathcal{O}(1/\Lambda^2,1/\Lambda^4)]$} & \multirow{2}{*}{(SM + $O_G$)/SM}  \\
\cline{1-6}
Subprocess	&	$\sigma$ [pb]	&	fraction [\%]	&	Subprocess	&	$\sigma$ [pb]	&	fraction [\%]	&		\\
\hline													
$gg \rightarrow gggg$	&	1.96$\cdot 10^{0}$	&	11.24	&	$gg \rightarrow gggg$	&	2.18$\cdot 10^{0}$	&	11.91	&	1.11	\\
$gg \rightarrow ggqq$	&	3.40$\cdot 10^{-1}$	&	1.95	&	$gg \rightarrow ggqq$	&	3.97$\cdot 10^{-1}$	&	2.17	&	1.17	\\
$gg \rightarrow qqqq$	&	7.10$\cdot 10^{-3}$	&	0.04	&	$gg \rightarrow qqqq$	&	8.58$\cdot 10^{-3}$	&	0.05	&	1.21	\\
$qg \rightarrow gggq$	&	8.19$\cdot 10^{0}$	&	46.97	&	$qg \rightarrow gggq$	&	8.57$\cdot 10^{0}$	&	46.91	&	1.05	\\
$qg \rightarrow gqqq$	&	9.13$\cdot 10^{-1}$	&	5.23	&	$qg \rightarrow gqqq$	&	9.82$\cdot 10^{-1}$	&	5.37	&	1.08	\\
$qq \rightarrow gggg$	&	4.61$\cdot 10^{-2}$	&	0.26	&	$qq \rightarrow gggg$	&	6.82$\cdot 10^{-2}$	&	0.37	&	1.48	\\
$qq \rightarrow ggqq$	&	5.75$\cdot 10^{0}$	&	32.94	&	$qq \rightarrow ggqq$	&	5.83$\cdot 10^{0}$	&	31.88	&	1.01	\\
$qq \rightarrow qqqq$	&	2.36$\cdot 10^{-1}$	&	1.35	&	$qq \rightarrow qqqq$	&	2.44$\cdot 10^{-1}$	&	1.33	&	1.03	\\
total	&	1.74$\cdot 10^{+1}$	&	100.00	&	total	&	1.83$\cdot 10^{+1}$	&	100.00	&	1.05	\\
\hline
\hline
\multicolumn{7}{c}{$S_T > 3$ TeV} \\
\hline
\multicolumn{3}{c ||}{SM} & \multicolumn{3}{c ||}{SM + $O_G [\mathcal{O}(1/\Lambda^2,1/\Lambda^4)]$} & \multirow{2}{*}{(SM + $O_G$)/SM}  \\
\cline{1-6}
Subprocess	&	$\sigma$ [pb]	&	fraction [\%]	&	Subprocess	&	$\sigma$ [pb]	&	fraction [\%]	&		\\
\hline													
$gg \rightarrow gggg$	&	5.10$\cdot 10^{-2}$	&	5.68	&	$gg \rightarrow gggg$	&	7.88$\cdot 10^{-2}$	&	7.34	&	1.54	\\
$gg \rightarrow ggqq$	&	8.13$\cdot 10^{-3}$	&	0.90	&	$gg \rightarrow ggqq$	&	1.45$\cdot 10^{-2}$	&	1.35	&	1.79	\\
$gg \rightarrow qqqq$	&	1.52$\cdot 10^{-4}$	&	0.02	&	$gg \rightarrow qqqq$	&	2.93$\cdot 10^{-4}$	&	0.03	&	1.93	\\
$qg \rightarrow gggq$	&	3.71$\cdot 10^{-1}$	&	41.22	&	$qg \rightarrow gggq$	&	4.58$\cdot 10^{-1}$	&	42.71	&	1.24	\\
$qg \rightarrow gqqq$	&	3.91$\cdot 10^{-2}$	&	4.35	&	$qg \rightarrow gqqq$	&	5.17$\cdot 10^{-2}$	&	4.82	&	1.32	\\
$qq \rightarrow gggg$	&	3.25$\cdot 10^{-3}$	&	0.36	&	$qq \rightarrow gggg$	&	1.01$\cdot 10^{-2}$	&	0.94	&	3.11	\\
$qq \rightarrow ggqq$	&	4.11$\cdot 10^{-1}$	&	45.70	&	$qq \rightarrow ggqq$	&	4.41$\cdot 10^{-1}$	&	41.14	&	1.07	\\
$qq \rightarrow qqqq$	&	1.60$\cdot 10^{-2}$	&	1.78	&	$qq \rightarrow qqqq$	&	1.79$\cdot 10^{-2}$	&	1.67	&	1.12	\\
total	&	8.99$\cdot 10^{-1}$	&	100.00	&	total	&	1.07$\cdot 10^{0}$	&	100.00	&	1.19	\\
\hline
\hline
\multicolumn{7}{c}{$S_T > 4$ TeV} \\
\hline
\multicolumn{3}{c ||}{SM} & \multicolumn{3}{c ||}{SM + $O_G [\mathcal{O}(1/\Lambda^2,1/\Lambda^4)]$} & \multirow{2}{*}{(SM + $O_G$)/SM}  \\
\cline{1-6}
Subprocess	&	$\sigma$ [pb]	&	fraction [\%]	&	Subprocess	&	$\sigma$ [pb]	&	fraction [\%]	&		\\
\hline												
$gg \rightarrow gggg$	&	1.75$\cdot 10^{-3}$	&	2.52	&	$gg \rightarrow gggg$	&	4.68$\cdot 10^{-3}$	&	4.54	&	2.67	\\
$gg \rightarrow ggqq$	&	2.67$\cdot 10^{-4}$	&	0.38	&	$gg \rightarrow ggqq$	&	9.04$\cdot 10^{-4}$	&	0.88	&	3.39	\\
$gg \rightarrow qqqq$	&	4.62$\cdot 10^{-6}$	&	0.01	&	$gg \rightarrow qqqq$	&	1.74$\cdot 10^{-5}$	&	0.02	&	3.77	\\
$qg \rightarrow gggq$	&	2.32$\cdot 10^{-2}$	&	33.40	&	$qg \rightarrow gggq$	&	3.99$\cdot 10^{-2}$	&	38.75	&	1.72	\\
$qg \rightarrow gqqq$	&	2.31$\cdot 10^{-3}$	&	3.32	&	$qg \rightarrow gqqq$	&	4.48$\cdot 10^{-3}$	&	4.35	&	1.94	\\
$qq \rightarrow gggg$	&	3.11$\cdot 10^{-4}$	&	0.45	&	$qq \rightarrow gggg$	&	2.28$\cdot 10^{-3}$	&	2.21	&	7.33	\\
$qq \rightarrow ggqq$	&	4.01$\cdot 10^{-2}$	&	57.73	&	$qq \rightarrow ggqq$	&	4.87$\cdot 10^{-2}$	&	47.29	&	1.21	\\
$qq \rightarrow qqqq$	&	1.53$\cdot 10^{-3}$	&	2.20	&	$qq \rightarrow qqqq$	&	2.01$\cdot 10^{-3}$	&	1.95	&	1.32	\\
total	&	6.95$\cdot 10^{-2}$	&	100.00	&	total	&	1.03$\cdot 10^{-1}$	&	100.00	&	1.48	\\
\hline
\end{tabular}
\end{center}
 \caption{\label{table:4-jet} Contributions from the different partonic subprocesses to the production of four partonic jets at LHC13 in the SM ($\mathcal{O}(1/\Lambda^0)$) and when allowing at most one $O_G$ insertion in the amplitudes ($\mathcal{O}(1/\Lambda^2)$ and $\mathcal{O}(1/\Lambda^4)$). }  
\end{table}

\section{One-loop $O_G$ contribution to dijet production}
\label{app:oneloop}
In Table \ref{tab:nlodi-jet} we show the one-loop contributions of $O_G$ to dijet production for different partonic channels and different cuts on the jet transverse momentum. 
\begin{table}[!h]
\renewcommand{\arraystretch}{1.}
\scriptsize
\begin{center}
\begin{tabular}{c | c c | c c | c c | c}
\hline
\hline
\multicolumn{8}{c}{$p p \rightarrow j j $, $p_T(j) > 100$ GeV} \\
\hline
 & \multicolumn{2}{c |}{SM} &  \multicolumn{2}{c |}{$O_G [\mathcal{O}(1/\Lambda^2)]$@1-loop} & \multicolumn{2}{c |}{$O_G [\mathcal{O}(1/\Lambda^4)]$} & \multirow{2}{*}{(SM + $O_G$)/SM}  \\
\cline{1-7}
Subprocess	&	$\sigma$ [pb]	&	fraction [\%]	&	$\sigma$ [pb]	&	fraction [\%]	& $\sigma$ [pb]  & 	fraction [\%]	&	\\
\hline													
$gg \rightarrow gg$	&	5.25$\cdot 10^{+5}$	&	46.06	&	-5.31$\cdot 10^{0}$	&	-28.02	&	2.12$\cdot 10^{+1}$	&	66.49	&	1.00	\\
$gg \rightarrow qq$	&	1.98$\cdot 10^{+4}$	&	1.74	&	5.59$\cdot 10^{0}$	&	29.52	&	1.96$\cdot 10^{0}$	&	6.14	&	1.00	\\
$qg \rightarrow qg$	&	5.02$\cdot 10^{+5}$	&	44.06	&	2.28$\cdot 10^{+1}$	&	120.46	&	8.48$\cdot 10^{0}$	&	26.59	&	1.00	\\
$qq \rightarrow gg$	&	1.33$\cdot 10^{+3}$	&	0.12	&	4.16$\cdot 10^{-1}$	&	2.20	&	2.52$\cdot 10^{-1}$	&	0.79	&	1.00	\\
$qq \rightarrow qq$	&	9.15$\cdot 10^{+4}$	&	8.02	&	-4.58$\cdot 10^{0}$	&	-24.17	&	0.00	&	0.00	&	1.00	\\
total	&	1.14$\cdot 10^{+6}$	&	100.00	&	1.89$\cdot 10^{+1}$	&	100.00	&	3.19$\cdot 10^{+1}$	&	100.00	&	1.00	\\
\hline
\hline
\multicolumn{7}{c}{$p_T(j) > 200$ GeV} \\
\hline
 & \multicolumn{2}{c |}{SM} &  \multicolumn{2}{c |}{$O_G [\mathcal{O}(1/\Lambda^2)]$@1-loop} & \multicolumn{2}{c |}{$O_G [\mathcal{O}(1/\Lambda^4)]$} & \multirow{2}{*}{(SM + $O_G$)/SM}  \\
\cline{1-7}
Subprocess	&	$\sigma$ [pb]	&	fraction [\%]	&	$\sigma$ [pb]	&	fraction [\%]	& $\sigma$ [pb]  & 	fraction [\%]	&	\\
\hline													
$gg \rightarrow gg$	&	2.02$\cdot 10^{+4}$	&	37.60	&	-8.71$\cdot 10^{-1}$	&	-29.15	&	9.70$\cdot 10^{0}$	&	61.05	&	1.00	\\
$gg \rightarrow qq$	&	8.09$\cdot 10^{+2}$	&	1.51	&	8.16$\cdot 10^{-1}$	&	27.34	&	9.52$\cdot 10^{-1}$	&	5.99	&	1.00	\\
$qg \rightarrow qg$	&	2.64$\cdot 10^{+4}$	&	49.26	&	4.00$\cdot 10^{0}$	&	133.94	&	5.05$\cdot 10^{0}$	&	31.82	&	1.00	\\
$qq \rightarrow gg$	&	9.69$\cdot 10^{+1}$	&	0.18	&	1.09$\cdot 10^{-1}$	&	3.64	&	1.80$\cdot 10^{-1}$	&	1.13	&	1.00	\\
$qq \rightarrow qq$	&	6.13$\cdot 10^{+3}$	&	11.45	&	-1.07$\cdot 10^{0}$	&	-35.76	&	0.00	&	0.00	&	1.00	\\
total	&	5.36$\cdot 10^{+4}$	&	100.00	&	2.99$\cdot 10^{0}$	&	100.00	&	1.59$\cdot 10^{+1}$	&	100.00	&	1.00	\\
\hline
\hline
\multicolumn{7}{c}{$p_T(j) > 300$ GeV} \\
\hline
 & \multicolumn{2}{c |}{SM} &  \multicolumn{2}{c |}{$O_G [\mathcal{O}(1/\Lambda^2)]$@1-loop} & \multicolumn{2}{c |}{$O_G [\mathcal{O}(1/\Lambda^4)]$} & \multirow{2}{*}{(SM + $O_G$)/SM}  \\
\cline{1-7}
Subprocess	&	$\sigma$ [pb]	&	fraction [\%]	&	$\sigma$ [pb]	&	fraction [\%]	& $\sigma$ [pb]  & 	fraction [\%]	&	\\
\hline													
$gg \rightarrow gg$	&	2.43$\cdot 10^{+3}$	&	31.66	&	-2.69$\cdot 10^{-1}$	&	-32.17	&	5.06$\cdot 10^{0}$	&	56.54	&	1.00	\\
$gg \rightarrow qq$	&	1.01$\cdot 10^{+2}$	&	1.31	&	2.29$\cdot 10^{-1}$	&	27.44	&	5.15$\cdot 10^{-1}$	&	5.76	&	1.01	\\
$qg \rightarrow qg$	&	4.00$\cdot 10^{+3}$	&	52.10	&	1.24$\cdot 10^{0}$	&	148.20	&	3.24$\cdot 10^{0}$	&	36.22	&	1.00	\\
$qq \rightarrow gg$	&	1.87$\cdot 10^{+1}$	&	0.24	&	4.45$\cdot 10^{-2}$	&	5.32	&	1.32$\cdot 10^{-1}$	&	1.48	&	1.01	\\
$qq \rightarrow qq$	&	1.13$\cdot 10^{+3}$	&	14.69	&	-4.08$\cdot 10^{-1}$	&	-48.78	&	0.00	&	0.00	&	1.00	\\
total	&	7.68$\cdot 10^{+3}$	&	100.00	&	8.36$\cdot 10^{-1}$	&	100.00	&	8.94$\cdot 10^{0}$	&	100.00	&	1.00	\\
\hline
\hline
\multicolumn{7}{c}{$p_T(j) > 500$ GeV} \\
\hline
 & \multicolumn{2}{c |}{SM} &  \multicolumn{2}{c |}{$O_G [\mathcal{O}(1/\Lambda^2)]$@1-loop} & \multicolumn{2}{c |}{$O_G [\mathcal{O}(1/\Lambda^4)]$} & \multirow{2}{*}{(SM + $O_G$)/SM}  \\
\cline{1-7}
Subprocess	&	$\sigma$ [pb]	&	fraction [\%]	&	$\sigma$ [pb]	&	fraction [\%]	& $\sigma$ [pb]  & 	fraction [\%]	&	\\
\hline												
$gg \rightarrow gg$	&	1.25$\cdot 10^{+2}$	&	23.27	&	-4.73$\cdot 10^{-2}$	&	-43.40	&	1.68$\cdot 10^{0}$	&	49.19	&	1.01	\\
$gg \rightarrow qq$	&	5.46$\cdot 10^{0}$	&	1.02	&	3.15$\cdot 10^{-2}$	&	28.91	&	1.79$\cdot 10^{-1}$	&	5.24	&	1.04	\\
$qg \rightarrow qg$	&	2.92$\cdot 10^{+2}$	&	54.34	&	2.17$\cdot 10^{-1}$	&	198.57	&	1.48$\cdot 10^{0}$	&	43.41	&	1.01	\\
$qq \rightarrow gg$	&	1.80$\cdot 10^{0}$	&	0.34	&	1.09$\cdot 10^{-2}$	&	9.97	&	7.36$\cdot 10^{-2}$	&	2.16	&	1.05	\\
$qq \rightarrow qq$	&	1.13$\cdot 10^{+2}$	&	21.04	&	-1.03$\cdot 10^{-1}$	&	-94.06	&	0.00	&	0.00	&	1.00	\\
total	&	5.37$\cdot 10^{+2}$	&	100.00	&	1.09$\cdot 10^{-1}$	&	100.00	&	3.41$\cdot 10^{0}$	&	100.00	&	1.01	\\
\hline
\hline
\multicolumn{7}{c}{$p_T(j) > 1$ TeV} \\
\hline
 & \multicolumn{2}{c |}{SM} &  \multicolumn{2}{c |}{$O_G [\mathcal{O}(1/\Lambda^2)]$@1-loop} & \multicolumn{2}{c |}{$O_G [\mathcal{O}(1/\Lambda^4)]$} & \multirow{2}{*}{(SM + $O_G$)/SM}  \\
\cline{1-7}
Subprocess	&	$\sigma$ [pb]	&	fraction [\%]	&	$\sigma$ [pb]	&	fraction [\%]	& $\sigma$ [pb]  & 	fraction [\%]	&	\\
\hline												
$gg \rightarrow gg$	&	9.86$\cdot 10^{-1}$	&	11.92	&	-1.82$\cdot 10^{-3}$	&	-341.74	&	1.69$\cdot 10^{-1}$	&	35.90	&	1.17	\\
$gg \rightarrow qq$	&	4.60$\cdot 10^{-2}$	&	0.56	&	9.89$\cdot 10^{-4}$	&	186.06	&	1.93$\cdot 10^{-2}$	&	4.09	&	1.44	\\
$qg \rightarrow qg$	&	4.24$\cdot 10^{0}$	&	51.19	&	1.03$\cdot 10^{-2}$	&	1932.63	&	2.66$\cdot 10^{-1}$	&	56.40	&	1.07	\\
$qq \rightarrow gg$	&	3.59$\cdot 10^{-2}$	&	0.43	&	7.99$\cdot 10^{-4}$	&	150.30	&	1.71$\cdot 10^{-2}$	&	3.62	&	1.50	\\
$qq \rightarrow qq$	&	2.97$\cdot 10^{0}$	&	35.91	&	-9.71$\cdot 10^{-3}$	&	-1827.25	&	0.00	&	0.00	&	1.00	\\
total	&	8.28$\cdot 10^{0}$	&	100.00	&	5.31$\cdot 10^{-4}$	&	100.00	&	4.71$\cdot 10^{-1}$	&	100.00	&	1.06	\\
\hline
\end{tabular}
\end{center}
 \caption{\label{tab:nlodi-jet} Contributions from the different partonic subprocesses of di-jet production at the LHC13 in the SM and with exactly one insertion of the $O_G$ operator in the amplitudes, leading to matrix element contributions of order $\mathcal{O}(1/\Lambda^4)$ at the tree level and $\mathcal{O}(1/\Lambda^2)$ at the one-loop level, shown separately here. We remind the reader that tree-level interference contributions with exactly one insertion of the $O_G$ operator are exactly zero.} 
\end{table}

\bibliographystyle{JHEP}
\bibliography{bib}
\end{document}